\documentclass[journal]{IEEEtran}
\usepackage{graphicx}
\usepackage{color}
\usepackage{placeins}
\usepackage{float}
\usepackage{tabularx,colortbl}
\usepackage{graphics, theorem, times, amsfonts, amsmath, amssymb, cite, verbatim}
\usepackage{multicol,multirow,booktabs}
\usepackage{tikz}
\usepackage{framed}
\usetikzlibrary{shapes,arrows}
\tikzstyle{phantom vertex} = [ ellipse, 
                               anchor = center, 
                               minimum height = 1*\unit, 
                               minimum width  = 1*\unit,
                               inner sep=0pt,
                               anchor=center]
\tikzstyle{red vertex}   = [black, fill = red!20,   phantom vertex, draw]
\tikzstyle{black vertex} = [black, fill = black!20, phantom vertex, draw]
\tikzstyle{blue vertex}  = [black, fill = blue!20,  phantom vertex, draw]
\tikzstyle{green vertex} = [black, fill = green!20,  phantom vertex, draw]
\tikzstyle{yellow vertex} = [black, fill = yellow!20,  phantom vertex, draw]
\tikzstyle{vertex}       = [draw, phantom vertex]

\tikzstyle{point} = [ellipse, inner sep=0pt, draw, fill=white, anchor = center,
                     minimum height = 0.05*\unit, minimum width  = 0.05*\unit]

\usepackage{tikz-3dplot} %for tikz-3dplot functionality
\usepackage{pgfplots}
\usepackage{xcolor}
\usepackage{needspace}

\usepackage{hyperref}
\input{mysymbol.sty}
\usepackage{todonotes}
\usepackage{amsmath}

\usepackage{epstopdf}
\graphicspath{{figures/}}

\newtheorem{definition}{\hspace{0pt}\bf Definition}
\newcommand{\vc}[1]{{\bf #1}}
\newcommand{\ma}[1]{{\bf #1}}

\begin{document}

%\title{A Signal-Processing View on Brain Graphs for Functional Neuroimaging}
\title{A Graph Signal Processing Perspective on\\ Functional Brain Imaging}

%%%%%%%%%%%%%%%%%%%%%%%%%%%%%%%%%%%%%%%%%%%%%%%%%%%%%%%%%%%%%%%%%%%%%
%%%   A   U   T   H   O   R   %%%%%%%%%%%%%%%%%%%%%%%%%%%%%%%%%%%
%%%%%%%%%%%%%%%%%%%%%%%%%%%%%%%%%%%%%%%%%%%%%%%%%%%%%%%%%%%%%%%%%%%%%
%
\author{Weiyu Huang$^*$, Thomas A. W. Bolton$^*$, John D. Medaglia, Danielle S. Bassett,\\ Alejandro Ribeiro, Dimitri Van De Ville, \emph{Senior Member}

\thanks{Copyright (c) 2017 IEEE. Personal use of this material is permitted. However, permission to use this material for any other purposes must be obtained from the IEEE by sending a request to pubs-permissions@ieee.org.\newline Supported by ARO W911NF1710438, the Bertarelli Foundation, the Center for Biomedical Imaging (CIBM), NIH DP5-OD021352, the NIDCR R01-DC014960, the Perelman School of Medicine, the John D. and Catherine T. Mac Arthur Foundation, the Alfred P. Sloan Foundation, and the ISI Foundation. The authors indicated with $^*$ contributed equally. Weiyu Huang and Alejandro Ribeiro are with the Department of Electrical \& System Engineering, University of Pennsylvania, Philadelphia PA, United States. Thomas AW Bolton and Dimitri Van De Ville are with the Institute of Bioengineering/Center for Neuroprosthetics, \'Ecole Polytechnique F\'ed\'erale de Lausanne (EPFL), Lausanne, Switzerland, and the Department of Radiology and Medical Informatics, University of Geneva, Geneva, Switzerland. John D. Medaglia is with the Department of Psychology, Drexel University, Philadelphia PA, United States, and the Department of Neurology, Perelman School of Medicine, University of Pennsylvania, Philadelphia PA, United States. Danielle S. Bassett is with the Department of Bioengineering and the Department of Electrical \& System Engineering, University of Pennsylvania, Philadelphia PA, United States. Corresponding authors: A.~Ribeiro (email: aribeiro@seas.upenn.edu) and D.~Van De Ville (e-mail: dimitri.vandeville@epfl.ch). Some of the results in Section \ref{sec:applications} are adapted or reproduced with permission from \cite{medaglia2016}; see \cite{medaglia2016} for detailed discussion of the implications of these findings for human cognition.}}

\maketitle

\begin{abstract} 
Modern neuroimaging techniques provide us with unique views on brain structure and function; i.e., how the brain is wired, and where and when activity takes place. Data acquired using these techniques can be analyzed in terms of its network structure to reveal organizing principles at the systems level. Graph representations are versatile models where nodes are associated to brain regions and edges to structural or functional connections. Structural graphs model neural pathways in white matter that are the anatomical backbone between regions. Functional graphs are built based on functional connectivity, which is a pairwise measure of statistical interdependency between activity traces of regions. Therefore, most research to date has focused on analyzing these graphs reflecting structure or function. 

Graph signal processing (GSP) is an emerging area of research where signals recorded at the nodes of the graph are studied atop the underlying graph structure. An increasing number of fundamental operations have been generalized to the graph setting, allowing to analyze the signals from a new viewpoint. Here, we review GSP for brain imaging data and discuss their potential to integrate brain structure, contained in the graph itself, with brain function, residing in the graph signals. We review how brain activity can be meaningfully filtered based on concepts of spectral modes derived from brain structure. We also derive other operations such as surrogate data generation or decompositions informed by cognitive systems. In sum, GSP offers a novel framework for the analysis of brain imaging data.  
\end{abstract}

% keywords
\begin{IEEEkeywords} Brain, neuroimaging, network models, graph signal processing, functional MRI
\end{IEEEkeywords}

\IEEEpeerreviewmaketitle

%%%%%%%%%%%%%%%%%%%%%%%%%%%%%%%%%%%%%%%%%%%%%%%%%%%%%%%%%%%%%%%%%%%%%
%%%   I   N   T   R   O   D   U   C   T   I   O   N   %%%%%%%%%%%%%%%
%%%%%%%%%%%%%%%%%%%%%%%%%%%%%%%%%%%%%%%%%%%%%%%%%%%%%%%%%%%%%%%%%%%%%
%
\section{Introduction}\label{sec_introduction}
\IEEEPARstart{A}{dvances} in neuroimaging techniques such as magnetic resonance imaging (MRI) have provided opportunities to measure human brain structure and function in a non-invasive manner~\cite{Mather.2013}. Diffusion-weighted MRI allows to measure major fiber tracts in white matter and thereby map the structural scaffold that supports neural communication. Functional MRI (fMRI) takes an indirect estimate of the brain approximately each second, in the form of blood oxygenation level-dependent (BOLD) signals. An emerging theme in computational neuroimaging is to study the brain at the systems level with such fundamental questions as how it supports coordinated cognition, learning, and consciousness.  

Shaped by evolution, the brain has evolved connectivity patterns that often look haphazard yet are crucial in cognitive processes. The apparent importance of these {\it connectomes}, has motivated the emergence of network neuroscience as a clearly defined field to study the relevance of network structure for cognitive function~\cite{bullmore2009, calhoun2009, Bassett17}. The fundamental components in network neuroscience are graph models \cite{Newman.2010} where nodes are associated to brain regions and edge weights are associated with the strength of the respective connections. This connectivity structure can be measured directly by counting fiber tracts in diffusion weighted MRI or can be inferred from fMRI BOLD measurements. In the latter case, networks are said to be functional and represent a measure of co-activation, e.g., the pairwise Pearson correlation between the activation time series of nodes. Functional connectivity networks do not necessarily represent physical connections although it has been observed that there is a strong basis of anatomical support for functional networks \cite{Sporns.2016}.

Connectomes, structural and functional alike, have been successfully analyzed utilizing a variety of tools from graph theory and network science~\cite{Newman.2010}. These analyses have uncovered a variety of measures that reflect organizational principles of brain networks such as the presence of communities where groups of regions are more strongly connected between each other than with other communities~\cite{Sporns.2016, van2011rich}. Network analysis has also been related to behavioral and clinical measures by statistical methods or machine learning tools to study development, behavior, and ability \cite{richiardi1301, adali2014, Bassett17a}.

As network neuroscience expands from understanding connectomes into understanding how connectomes and functional brain activity support behavior, the study of {\it dynamics} has taken center stage. In addition, there is a rise of interest in analyzing and understanding dynamics of functional signals and with them, network structure. Such changes happen at different timescales, from years -- e.g., in developmental studies~\cite{Dosenbach.2010} -- down to seconds within a single fMRI run of several minutes~\cite{preti1701p}, or following tasks such as learning paradigms~\cite{Bassett17a, medaglia2016, huang2016}. So far, common approaches include examining changes in network structure (e.g., reflecting segregation and integration)~\cite{Sizemore.2017} or investigating time-resolved measures of the underlying functional signals~\cite{calhoun2014, Keilholz.2017,Karahanoglu.2017}. In the case of developmental studies, the evolution of structural networks is important, but large-scale anatomical changes do not occur in the shorter time scales that are involved in behavior and ability studies. In the latter case, the notion of a dynamic network itself makes little sense and the more pertinent objects of interest are the dynamic changes in brain activity signals \cite{medaglia2016, huang2016}. Inasmuch as brain activity is mediated by physical connections, the underlying network structure must be taken into account when studying these signals. Tools from the emerging field of graph signal processing (GSP) are tailored for this purpose.

Put simply, GSP addresses the problem of analyzing and extracting information from data defined not in regular domains such as time or space, but on more irregular domains that can be conveniently represented by a graph. The fundamental GSP concepts that we utilize to analyze brain signals are the graph Fourier transform (GFT) and the corresponding notions of graph frequency components and graph filters. These concepts are generalizations of the Fourier transform, frequency components, and filters that have been used in regular domains such as time and spatial grids \cite{sandryhaila2013, sandryhaila2014, shuman2013}. As such, they permit the decomposition of a graph signal into pieces that represent different levels of variability. We can define low graph frequency components representing signals that change slowly with respect to brain networks, and high graph frequency components representing signals that change swiftly with respect to the connectivity networks. This is crucial because low and high {\it temporal} variability have proven to be important in the analysis of neurological disease and behavior \cite{garrett2012, Heisz2012}. GFT-based decompositions permit a similar analysis of variability across regions of the brain for a fixed time -- a sort of {\it spatial} variability measured with respect to the connectivity pattern. We review a recent study \cite{medaglia2016} that such a decomposition can be used to explain individual cognitive differences, as illustrated in Figures~\ref{fig_switching_results} and~\ref{fig_correlation_by_region}, and offer other perspectives to apply graph signal processing to functional brain analytics. The theory of GSP has been growing rapidly in recent years, with development in areas including sampling theory~\cite{marques2016, chen2015}, stationarity~\cite{perraudin2016, marques2017} and uncertainty~\cite{Agaskar.2013,pasdeloup2015,Tsitsvero.2016,Teke.2017}, filtering~\cite{rabbat2016, tremblay2016, kotzagiannidis2016a}, directed graphs \cite{shafipour2017a}, and dictionary learning~\cite{chen2016}. Applications have been spanning many areas including neuroscience \cite{huang2016, liu2016}, imaging \cite{pang2016, thanou2016}, medical imaging \cite{kotzagiannidis2016}, video \cite{wang2014}, online learning \cite{shafipour2017}, and rating prediction \cite{kalofolias2014, huang2017a, ma2016diffusion}.

%GSP deals with how to transpose conventional signal processing operations to graph signals; i.e., signals that are defined on a graph and take a value at each node. At the core of many of these methods, the eigendecomposition of a graph shift operator is used to define a graph Fourier transform, which can then be exploited to define graph equivalents of operations such as filtering, sampling, translation, modulation, or randomization~\cite{shuman2013}. The prominent role of the graph spectral domain should come at no surprise given its central place in graph analysis~\cite{von2007tutorial}, going back to the Fiedler vector for graph clustering~\cite{Fiedler.1989}; however, GSP goes one step further by analyzing the graph signals, and not just the graph itself. 

In this work, we broadly cover how GSP can be applied for an elegant and principled analysis of brain activity. In Section~\ref{sec_graphs_and_signals}, we start by constructing a graph from structural connectivity---the backbone of the brain---and considering brain activity as graph signals. Then, in Section~\ref{sec:GSP}, we derive the graph spectral domain by the eigendecomposition of a graph shift operator. Such eigenmodes have already been recognized as useful by providing robust representation of the connectome in health and disease ~\cite{Wang.2017}. We introduce a number of graph signal operations that are particularly useful for processing the activity time courses measured at the nodes of the graph; i.e., filtering in terms of anatomically-aligned or -liberal modes, randomization preserving anatomical smoothness, and localized decompositions that can incorporate additional domain knowledge. 
In the following sections, we review a recent study in \cite{medaglia2016} demonstrating the relevance of these GSP tools as an integrated framework to consider structure and function: in the context of an attention task, we discuss the potential of GSP operations to capture cognitively relevant brain properties (Section~\ref{sec:applications}). We also provide avenues for utilizing GSP tools in the structure-informed study of functional brain dynamics (Section~\ref{sec:dynamics}), through the extraction of significant excursions in a particular structure/function regime (Section~\ref{subsec:excursions}), and by more elaborate uses of GSP building blocks that can broaden the analysis to the temporal frequency domain, or narrow it down to a localized subset of selected regions (Section~\ref{subsec:further}). As certain parts of the paper include specific neuroscience terminology, a summarizing table (Table \ref{tab_terminology}) is provided for the reader's reference.

%%%%%%%%%%%%%%%%%%%%%%%%%%%%%%%%%%%%%%%%%%%%%%%%%%%%%%%%%%%%%%%%%%%%%
%%%   T   A   B   L   E   %%%%%%%%%%%%%%%%%%%%%%%%%%%%%%%%%%%%%%%
%%%%%%%%%%%%%%%%%%%%%%%%%%%%%%%%%%%%%%%%%%%%%%%%%%%%%%%%%%%%%%%%%%%%%
%
\begin{table*}[t]
\caption{Neuroscience terminology used in the paper\vspace{-0mm}}
\begin{center}{
\begin{tabular}{cl}        \toprule  
  Terminology & \multicolumn{1}{c}{Meaning} \\\midrule
  fMRI &  Measurement of brain activity by detecting changes associated with blood flow \\
  BOLD effect & When neuronal activity is increased in a brain region, there is an increased amount of cerebral blood flow to the area \\
  Connectomes & A comprehensive map of neural connections in the brain \\
  Navon switching task & A task where participants are asked to switch their attention between global and local features  \\\midrule
  Subcortical system & Situated beneath the cerebral cortex \\
  Fronto-parietal & Implicated in executive functions such as cognitive control and working memory, among others\\
  Auditory system & Responsible for the sense of hearing, and involved in linguistic processing \\
  Cingulo-opercular & Implicated in cognitive control \\
  Somatosensory & Processes sensations, or external stimuli, from our environment \\
  Ventral/dorsal attention & Typically reorient attention towards the salient stimuli, and respond with activation increases, respectively \\
  Default mode  & Involved in processing of one's self, thinking about others, remembering the past, and thinking about the future \\
  Visual system & Enables the ability to process visual detail \\\bottomrule
\end{tabular}\vspace{-4mm}}
\label{tab_terminology}
\end{center}
\end{table*}

%%%%%%%%%%%%%%%%%%%%%%%%%%%%%%%%%%%%%%%%%%%%%%%%%%%%%%%%%%%%%%%%%%%%%
%%%   S   E   C   T   I   O   N   %%%%%%%%%%%%%%%%%%%%%%%%%%%%%%%%%%%
%%%%%%%%%%%%%%%%%%%%%%%%%%%%%%%%%%%%%%%%%%%%%%%%%%%%%%%%%%%%%%%%%%%%%
%
\section{Brain Graphs and Brain Signals}
\label{sec_graphs_and_signals}

\begin{figure*}
    \begin{framed}
    \begin{center}\sc Callout 1: Estimating Brain Graphs.\end{center} \small
    MRI allows the acquisition of detailed structural information about the brain. The brain graph investigated in the present article was acquired on a Siemens 3.0T Tim Trio with a T1-weighted anatomical scan. Twenty-eight healthy individuals volunteered for the experiment. We followed a parallel strategy for data acquisition and construction of streamline adjacency matrices as in \cite{Gu2015}. First, DSI scans sampled 257 directions using a Q5 half-shell acquisition scheme with a maximum $b$-value of 5,000 and an isotropic voxel size of 2.4 mm. We utilized an axial acquisition with repetition time (TR) = 5 s, echo time (TE)= 138 ms, 52 slices, field of view (FoV) (231, 231, 125 mm). We acquired a three-dimensional SPGR T1 volume (TE = minimal full; flip angle = 15 degrees; FOV = 24 cm) for anatomical reconstruction. 
     Second, diffusion spectrum imaging (DSI) was performed to establish structural connectivity. DSI data were reconstructed in DSI Studio %\footnote{www.dsi-studio.labsolver.org} 
    using $q$-space diffeomorphic reconstruction (QSDR)\cite{yeh2011estimation}. QSDR computes the quantitative anisotropy in each voxel, which is used to warp the brain to a template QA volume in Montreal Neurological Institute (MNI) space. Then, spin density functions were again reconstructed with a mean diffusion distance of 1.25 mm using three fiber orientations per voxel. Fiber tracking was performed in DSI studio with an angular cutoff of 35$^\circ$, step size of 1.0 mm, minimum length of 10 mm, spin density function smoothing of 0.0, maximum length of 400 mm, and a QA threshold determined by diffusion-weighted imaging (DWI) signal in the colony-stimulating factor. Deterministic fiber tracking using a modified FACT algorithm was performed until 1,000,000 streamlines were reconstructed for each individual.
    Third, each anatomical scan was segmented using FreeSurfer\cite{fischl2012freesurfer}, and parcellated using the connectome mapping toolkit \cite{cammoun2012mapping}. A parcellation scheme including $N=87$ regions was registered to the B0 volume from each subject's DSI data. The B0 to MNI voxel mapping produced via QSDR was used to map region labels from native space to MNI coordinates. To extend region labels through the grey-white matter interface, the atlas was dilated by 4mm \cite{cieslak2014}. We used FSL to nonlinearly register the individual T1 scans to MNI space. 
    By combining parcellation and streamline information, we constructed subject-specific structural connectivity matrices, whose elements represent the number of streamlines connecting two different regions \cite{Hermundstad2013}, divided by the sum of their volumes \cite{hagmann2008mapping}. This process yields the weighted adjacency matrix $\bbA \in \reals^{N\times N}$ for each individual considered here.
    \end{framed}

	\centerline{\includegraphics[width=\textwidth]{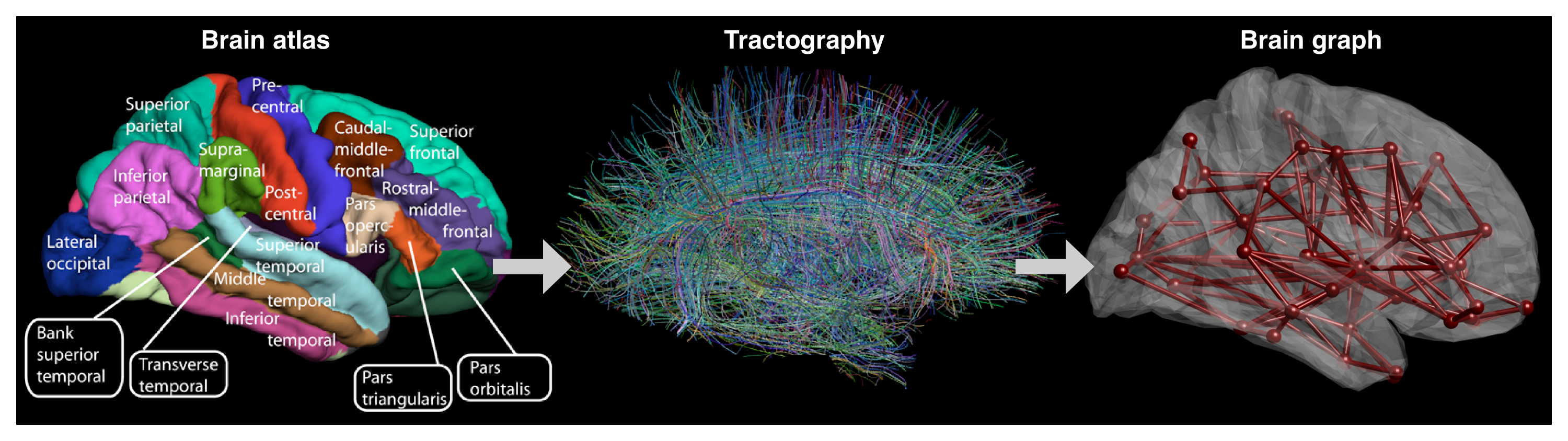}}
	\caption{\textbf{Estimating brain graphs}. Knowledge from an anatomical atlas based on anatomical features such as gyri and sulci (left) is combined with MRI structural connectivity extracted from diffusion-weighted MRI (middle), which can then be used to estimate the brain graph (right). [Adapted from~\cite{hagmann2008mapping}].}
	\label{fig:brain-graph}
\end{figure*}

Brain networks describe physical connection patterns between brain regions. These connections are mathematically described by a weighted graph $\ccalG := (\ccalV, \bbA)$ where $\ccalV = \{1, 2, \dots, N\}$ is a set of $N$ nodes associated with specific brain regions and $\bbA\in\reals_+^{N\times N}$ is a weighted adjacency matrix with entries $A_{i,j}$ representing the strength of the physical connection between brain regions $i$ and $j$. Some readers may prefer to consider the graph as a tuple $\ccalG = (\ccalV, \ccalE)$ where $\ccalE \subset \ccalV \times \ccalV$ describes the existence of physical connections between pairs of brain regions; each edge $(i, j) \in \ccalE$ has an underlying weight $A_{i,j}$ quantifying the strength of the connection. In this paper, we use $\ccalG := (\ccalV, \bbA)$ because it is more concise; notice that we can infer the existence of an edge $(i, j) \in \ccalE$ from the weight in the adjacency matrix if $A_{i, j} > 0$.

The brain regions encoded in the nodes of $\ccalV$ are macro-scale parcels of the brain that our current understanding of neuroscience deems anatomically or functionally differentiated. There are various parcellations in use in the literature that differ mostly in their level of resolution \cite{Zalesky_atlasing,sporns2011}. As an example, the networks that we study here consist of $N=82$ regions from the Desikan-Killiany anatomical atlas \cite{desikan2006automated} combined with the Harvard-Oxford subcortical parcels \cite{kennedy1998gyri}. A schematic representation of a few labeled brain regions is shown in Figure~\ref{fig:brain-graph} (left). 

\begin{figure*}[t]
    \begin{framed}
    \begin{center}{\sc Callout 2: Estimating Brain Signals.}\end{center}\small
    To derive the studied brain activity signals, functional MRI (fMRI) runs were acquired during the same scanning sessions as the DSI data on a 3.0T Siemens Tim Trio whole-body scanner with a whole-head elliptical coil by means of a single-shot gradient-echo T2* (TR = 1500 ms; TE = 30 ms; flip angle = 60$^\circ$; FOV = 19.2 cm, resolution 3mm x 3mm x 3mm). Preprocessing was performed using FEAT \cite{jenkinson2012fsl}, and included skull-stripping with BET \cite{Smith2000} to remove non-brain material, motion correction with MCFLIRT \cite{jenkinson2012fsl}, slice timing correction (interleaved), spatial smoothing with a 6mm 3D Gaussian kernel, and high-pass temporal filtering to reduce low-frequency artifacts. We also performed EPI unwrapping with fieldmaps in order to improve subject registration to standard space. Native image transformation to a standard template was completed using FSL's affine registration tool, FLIRT \cite{jenkinson2012fsl}. 
    Subject-specific functional images were co-registered to their corresponding high-resolution anatomical images via a Boundary Based Registration technique \cite{greve2009accurate} and were then registered to the standard MNI-152 structural template via a 12-parameter linear transformation. 
    Finally, we extracted region-averaged BOLD signals using the same atlas as for the structural analysis. At the end of this pipeline, we are thus left with a signal matrix $\bbX \in \reals^{N\times T}$ for each subject, reflecting the activity levels of all brain regions over time.
    \end{framed}

	\centerline{\includegraphics[width=1 \textwidth, trim=0cm 0.0cm 0cm 0.0cm, clip=true]{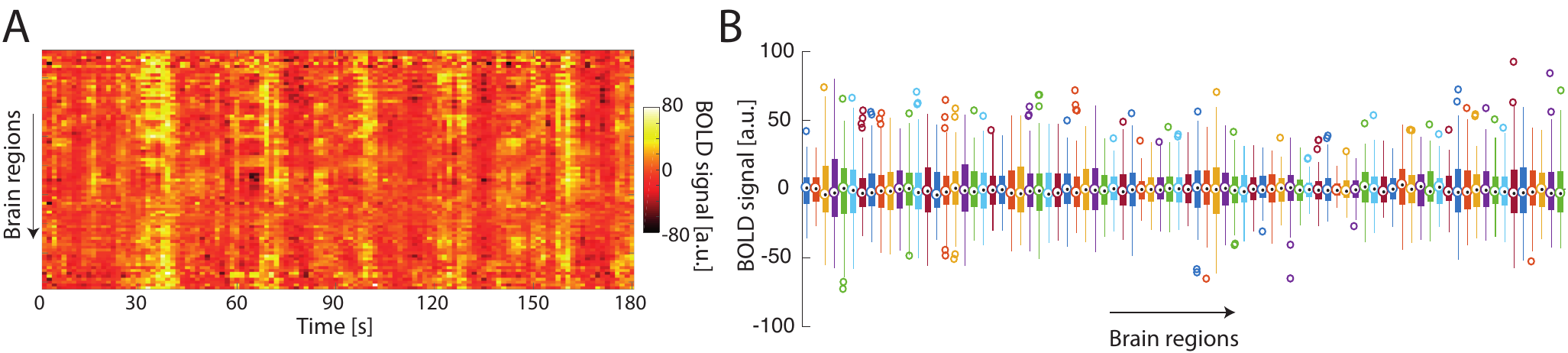}}
	\caption{\textbf{Example brain activity signals}. (\textbf{A}) For an example subject, the heat map of BOLD magnitude activity across brain regions (vertically) and time points (horizontally). Brain activity signals can be considered as a two-dimensional matrix, indexed in both the temporal and spatial domains. From the temporal perspective, there are certain time instances (e.g., in this case, between 30 and 40 seconds and between 70 and 80 seconds) when BOLD magnitudes are in general stronger than for others. From the spatial perspective, signals on most brain regions change in the same direction, but there are certain brain regions where their changes do not follow the main trend. As we will see, low and high graph frequency components, respectively, can be used to extract these two different pieces of information. (\textbf{B}) For the same subject, distribution of fMRI BOLD values for each brain region (horizontally) across all time points. Different brain regions exhibit different levels of variability, but in general, the wide variance of BOLD signals complicates data analysis. For each brain region, edges of the box denote 25th and 75th percentiles respectively; whiskers extend to the extreme points not considered to be outliers; circles denote outliers, which are values beyond 1.5 times the interquartile range away from the edges of the box.  \vspace{-4mm}}
	\label{fig:brain-signals}
\end{figure*}

The entries $A_{ij}$ of the adjacency matrix $\bbA$ measure the strength of the axonal connection between region $i$ and region $j$. This strength is a simple count of the number of streamlines that connect the regions, and can be estimated with diffusion spectrum imaging (DSI) \cite{Gu2015} --- see Figure~\ref{fig:brain-graph} for an illustration of the pipeline and Callout 1 for details on the specific techniques that are used for this purpose. In a situation of healthy development and an absence of trauma, nodes in brain graphs are the same across individuals. 
%Connections between brain regions, however, exhibit significant variability from person to person.  % I would not claim this too strongly because we use an average structural connectome later on...
Inter-subject variability of structural connectivity has demonstrated clinical value as it has been reliably associated with neurological \cite{medaglia2017brain, braun2015} and psychological \cite{gaetz2014} disorders. 

Besides structural connectivity, it is also possible to acquire brain activity signals $\bbx\in\reals^N$ such that the value of the $i$\textsuperscript{th} component $x_i$ quantifies neuronal activity in brain region $i$ --- see Figure~\ref{fig:brain-signals} for an illustration of these BOLD signals and Callout 2 for details on the methods. BOLD signals for all the $N$ studied brain regions are acquired over $T$ successive time points, and therefore, we define the matrix $\bbX\in\reals^{N\times T}$ such that its $j$\textsuperscript{th} column codifies brain activity at time $j$. An example of such a brain signal matrix is provided in Figure~\ref{fig:brain-signals}A, with the corresponding distribution of values for each brain region illustrated in Figure~\ref{fig:brain-signals}B.

Brain activity signals carry dynamic information that is not only useful for the study of pathology \cite{braun2015,Tagliazucchi2010,Christoff2016}, but also enables us to gain insight into human cognitive abilities \cite{van2009efficiency,haken2013principles,garrett2013}. Whereas physical connectivity can be seen as a long-term property of individuals that changes slowly over the course of years, brain activity signals display meaningful fluctuations at second or sub-second time scales that reflect how different parts of the brain exchange and process information in the absence of any external stimulus, and how they are recruited to meet emerging cognitive challenges. There is increasing evidence that differences in activation patterns across individuals tightly relate to behavioral variability \cite{bassett2011dynamic, Thompson2013, sporns2014contributions, huang2016}.

To the extent that brain activity signals are generated on top of the physical connectivity substrate, brain graphs and brain signals carry complementary information and should be studied in conjunction. This has been a challenge in neuroscience due to the unavailability of appropriate methods for performing this joint analysis. Here, we advocate for the use of GSP tools, as detailed in the following section.

%%%%%%%%%%%%%%%%%%%%%%%%%%%%%%%%%%%%%%%%%%%%%%%%%%%%%%%%%%%%%%%%%%%%%
%%%   F   I   G   U   R   E   %%%%%%%%%%%%%%%%%%%%%%%%%%%%%%%%%%%
%%%%%%%%%%%%%%%%%%%%%%%%%%%%%%%%%%%%%%%%%%%%%%%%%%%%%%%%%%%%%%%%%%%%%
%
\begin{figure*}
   \includegraphics[width=\textwidth]{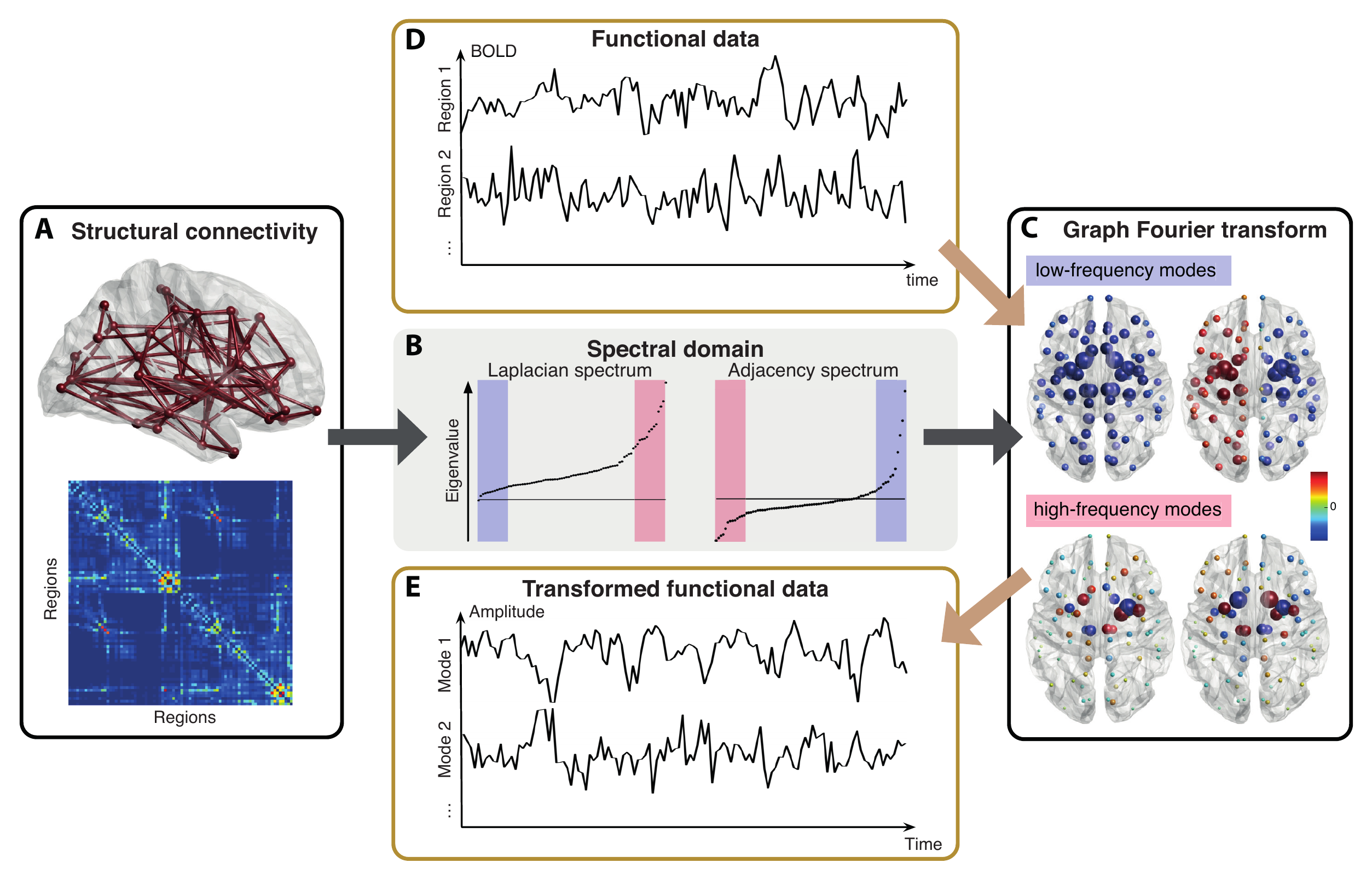}
   \caption{\label{fig:overview} \textbf{Graph signal processing for brain imaging}. (\textbf{A}) Structural connectivity from diffusion-weighted MRI, as seen in the form of a sagittal brain view (top) or of an adjacency matrix where the weights represent the strength of the structural connections (bottom), is used to build a graph representing the brain's wiring scaffold. (\textbf{B}) Through the eigendecomposition of the Laplacian (left plot) or adjacency (right plot) matrix, this structural graph can be analyzed in the spectral domain. The smallest Laplacian eigenvalues (or most positive adjacency eigenvalues) (labeled in blue) are associated with low-frequency modes on the graph (\textbf{C}, top brain views), while the largest Laplacian eigenvalues (or most negative adjacency eigenvalues) (labeled in red) are associated with high-frequency modes (\textbf{C}, bottom brain views). Together, these modes define the graph Fourier transform. Functional MRI data measured at the nodes of the graph (\textbf{D}) can be decomposed using these modes, and transformed by means of graph signal processing tools (\textbf{E}).}
\end{figure*}

%%%%%%%%%%%%%%%%%%%%%%%%%%%%%%%%%%%%%%%%%%%%%%%%%%%%%%%%%%%%%%%%%%%%%
%%%   S   E   C   T   I   O   N   %%%%%%%%%%%%%%%%%%%%%%%%%%%%%%%%%%%
%%%%%%%%%%%%%%%%%%%%%%%%%%%%%%%%%%%%%%%%%%%%%%%%%%%%%%%%%%%%%%%%%%%%%
%
\section{Graph Signal Processing for Neuroimaging}
\label{sec:GSP}

The GSP perspective is to interpret the brain signal $\bbx$ as a graph signal that is supported on the brain graph $\ccalG= (\ccalV,\bbA)$. Here, we introduce the fundamental operations that we will need for processing neuroimaging data in a meaningful way.

\subsection{Graph Fourier Transform}
\label{subsec:GFT}
The focus of GSP is not on analyzing the brain graph $\mathcal{G}$ \emph{per se}, but on using that graph to analyze brain signals $\bbx$. For a graph with positive edge weights, we consider a \emph{graph shift operator} that captures the connectivity pattern of $\ccalG$; we can choose the adjacency matrix $\bbA$ \cite{sandryhaila2014,sandryhaila2013} or the graph Laplacian $\bbL = \bbD - \bbA$\cite{shuman2013, Chung97}, where the degree matrix $\ma D$ contains the degree of each node on its diagonal: $D_{i,i} = \sum_{j \in \ccalV}A_{i,j}$. There are also several variants of the graph Laplacian~\cite{von2007tutorial} such as the symmetric normalized graph Laplacian $\ma{L}_\text{sym}=\ma{D}^{-1/2}\ma{L}\ma{D}^{-1/2}$ that factors out differences in degree and is thus only reflecting relative connectivity, or the random-walk normalized graph Laplacian: $\ma{L}_\text{rw}=\ma{D}^{-1}\ma{L}$. Generalizations of the graph Laplacian also exist for graphs with negative weights \cite{ma2016diffusion, kunegis2010spectral}. 

Let us denote the graph shift operator as $\ma S$ and assume henceforth that $\bbS$ is diagonalizable using singular value decomposition or Jordan decomposition, so that $\bbS=\bbV\bbLambda\bbV^{-1}$ where $\bbLambda$ is a diagonal matrix containing the eigenvalues $\lambda_k\in\mathbb{C}$, $k=0,\ldots,N-1$, and $\bbV = [\bbv_0,\bbv_1,\ldots,\bbv_{N-1}]$. When $\bbS$ is symmetric, we have that $\bbV$ is real and unitary, which implies $\bbV^{-1}=\bbV^\top$. The intuition behind examining $\bbS$ as an operator is to represent a transformation that characterizes exchanges between neighboring nodes. The eigendecomposition of $\bbS$ is then used to define the graph spectral domain.

%%%%%%%%%%%%%%%%%%%%%%%%%%%%%%%%%%%%%%%%%%%%%%%%%%%%%%%%%%%%%%%%%%%%%
%%%   D   E   F   I   N   I   T   I   O   N   %%%%%%%%%%%%%%%%%%%%%%%%%%%%%%%%%%%
%%%%%%%%%%%%%%%%%%%%%%%%%%%%%%%%%%%%%%%%%%%%%%%%%%%%%%%%%%%%%%%%%%%%%
%
\begin{definition}\label{def_GFT} 
Consider a signal $\bbx\in\reals^N$ and a graph shift operator $\bbS=\bbV\bbLambda\bbV^{-1}\in\reals^{N\times N}$. Then, the vectors
	\begin{align}\label{eqn_GFT}
	\tbx = \bbV^\top \bbx \hspace{.5cm}\text{and}\hspace{.5cm} \bbx = \bbV \tbx
	\end{align}
	form a Graph Fourier Transform (GFT) pair \cite{sandryhaila2013,shuman2013}. 
\end{definition}

%There are several reasons that justify the association of the GFT with the Fourier transform. Mathematically, it is just a matter of definition that if the vectors $\bbv_k$ are of the form $\bbv_k = [1, e^{j2\pi k/n}, \ldots, e^{j2\pi k(n-1)/n}]^\top$, the GFT and iGFT in Definition \ref{def_GFT} reduce to the conventional time domain Fourier and inverse Fourier transforms. More deeply, it is not difficult to see that if the graph $\ccalG$ is a cycle, the vectors $\bbv_k$ in are of the form $\bbv_k = [1, e^{j2\pi k/n}, \ldots, e^{j2\pi k(n-1)/n}]^\top$. Since cycle graphs are representations of discrete periodic signals, it follows that the GFT of a time signal is equivalent to the conventional discrete Fourier transform.

%%%%%%%%%%%%%%%%%%%%%%%%%%%%%%%%%%%%%%%%%%%%%%%%%%%%%%%%%%%%%%%%%%%%%
%%%   M   A   I   N      M   A   T   T   E   R   %%%%%%%%%%%%%%%%%%%%%%%%%%%%%%%%%%%
%%%%%%%%%%%%%%%%%%%%%%%%%%%%%%%%%%%%%%%%%%%%%%%%%%%%%%%%%%%%%%%%%%%%%
%

The GFT encodes the notion of variability for graph signals akin to the one that the Fourier transform encodes for temporal signals. When choosing the adjacency matrix $\ma A$ as a shift operator for directed graphs~\cite{sandryhaila2013, sandryhaila2014,Mhaskar.2017}, the eigenvalues $\lambda_k$ can be complex; the smaller the distance between $\lambda_k$ and $|\lambda_{\max}(\bbS)|$ in the complex spectrum, the lower the frequency it represents. This idea is based on defining the total variation of a graph signal $\bbx$ as $\| \bbx - \bbS \bbx / \lambda_{\max}(\bbS) \|_1$, with smoothness being associated to small values of total variation.  Then, given a $(\lambda_k,\bbv_k)$ pair, one has total variation with $\| 1-\lambda_k/\lambda_{\max}(\bbS)\|_1 \|\bbv_k\|_1$, which provides an intuitive way to order the different frequencies. Graph frequency ordering becomes more obvious for undirected graphs and thus symmetric adjacency matrices, as eigenvalues become real numbers. Specifically, the quadratic form of $\bbA$ is given by $\lambda_k = \bbv_k^\top \bbA \bbv_k = \sum_{i \neq j} A_{i,j} [\bbv_k]_i [\bbv_k]_j$. In this setting, lower frequencies will be associated to larger eigenvalues, to represent the fact that highly connected nodes in the graph possess signals with the same sign and similar values. 

When using the graph Laplacian $\bbL$ as a shift operator~\cite{shuman2013} for an undirected graph, the quadratic form of $\bbL$ is given by $\lambda_k = \bbv_k^\top \bbL \bbv_k = \sum_{i\neq j} A_{i,j} ([\bbv_k]_i - [\bbv_k]_j)^2$. If the signal variations follow the graph structure, the resulting value will be low. Thus, in this setup, the eigenvectors associated to smaller eigenvalues can be regarded as the graph lower frequencies. Further, the basis $\bbV$ is then a common solution to several well known signal processing problems, including Laplacian embedding, where the aim is to find a mapping of the graph nodes on a line so that connected nodes stay as close as possible, or in other words, to minimize $\bbx^\top \bbL \bbx$ under the constraints $\bbx^\top \bbx = 1$ and $\bbx^\top \mathbf{1} = 0$ \cite{Belkin2003}. Another is the classical graph cut problem \cite{Shi2000,Newman2013}, where the goal is to partition a graph into sub-communities of nodes with as few cross-connections as possible, with a similar obtained solution upon relaxation of the $x_i = \pm 1$ constraint.

Besides a decomposition along the spatial domain, we can also use the classical discrete Fourier transform (DFT) to decompose $\bbX$ along its temporal dimension as:
\begin{align}\label{eqn_Fourier_representation}
	\hat{\bbX} = \bbX \bbF^{H},
\end{align}
where $\cdot^{H}$ indicates the Hermitian transpose, and $\bbF$ is the Fourier matrix. $\hat{\bbX} \in \mathbb{C}^{N\times T}$ contains $T$ Fourier coefficients for each of the $N$ time courses. Filtering can then be applied by multiplying with a diagonal matrix $\bbH$ defined by the windowing function $[\bbH]_{i,i}=h(\lambda_i)$, with the filtered output given by:
\begin{align}\label{eqn_Fourier_filtered_signal}
	\bbY_{\bbH} = \bbX \bbF^{H} \bbH \bbF.
\end{align}

Notice that the DFT can also be obtained using the graph formalism by considering cycle graphs that represent discrete periodic signals~\cite{leonardi1302,shuman2013,marques2016, sandryhaila2014}. Specifically, we consider the undirected graph $\ccalG$ with adjacency matrix $\bbA_\text{cycle}$ such that $[\bbA_\text{cycle}]_{i, i+1 \mod T} = [\bbA_\text{cycle}]_{i,i-1 \mod T}=1$, and $[\bbA_\text{cycle}]_{i, j} = 0$ otherwise. For this graph, the eigenvectors of its adjacency $\bbA_\text{cycle}$ or its Laplacian matrix $\bbL_\text{cycle} = 2\bbI - \bbA_\text{cycle}$ satisfy $\bbV = \bbF$. Since cycle graphs are representations of discrete periodic signals, it follows that the GFT of a time signal is equivalent to the conventional discrete Fourier transform. In other words, a GFT is equivalent to a DFT for cyclic graphs. We also note that it is possible to combine DFT and GFT to investigate the joint spatial-temporal frequency, i.e., $\hat{\tilde{\bbX}} = \bbV^\top \bbX \bbF^H$. Such analytical efforts have been developing recently; see \cite{perraudin2016, perraudin2017, grassi2017, sandryhaila2014a} for more details.
 
\subsection{Graph Signal Filtering}
\label{subsec:graph_filtering}
Given the above relationships, it becomes possible to manipulate the graph signals stored in the matrix $\bbX$ by extracting signal components associated to different graph frequency ranges. Specifically, we can define the diagonal filtering matrix $\bbG$, where $[\bbG]_{i,i} = g(\lambda_i)$ is the frequency response for the graph frequency associated with eigenvalue $\lambda_i$, and retrieve the filtered signals as:
\begin{align}\label{eqn_GFT_lowpass}
	\bbY_{\bbG} = \bbV \bbG \bbV^\top \bbX.
\end{align}
Generic filtering operations can now be defined for the graph setting, such as ideal low-pass filtering, where $g(\lambda_i)$ would be 1 for $\lambda_i$ corresponding to low-frequency modes, and $0$ otherwise. 

Using the definition of the GFT pair, the effect of the filtering in \eqref{eqn_GFT_lowpass} on the graph spectral coefficients is directly visible from $\tilde{\ma Y}_{\ma G}=\ma G \tilde{\ma X}$. This also allows to generalize the convolution operation of a graph signal $\vc x$ by a filter defined through the spectral window $g$ as~\cite{shuman2013}:
$$
   [\vc y_{\ma G}]_{k'}=\sum_{k=0}^{N-1} [\vc v_k]_{k'} g(\lambda_k) \tilde{x}_k. 
$$
It is also possible to translate the operation to the vertex domain by considering the Taylor approximation of the window function $g(\lambda)=\sum_{m=0}^{M} c_m \lambda^m$:
$$
  \ma Y_{\ma G}=\sum_{m=0}^{M} c_m \ma S^m \ma X, 
$$
which uses iterated versions of the shift operator $\ma S$. Other operations such as translation, modulation, or dilation can be generalized in a similar way~\cite{shuman2013}.

\subsection{Generation of Graph Surrogate Signals}
\label{subsec:SpectralSurrogate}

A pivotal aspect in any research field is to assess the significance of obtained results through statistical testing. More precisely, one aims to invalidate the \textit{null hypothesis}, which expresses the absence of the effect of interest. Standard parametric tests such as the well-known $t$-test assume independent and identically distributed Gaussian noise, which makes a weak null hypothesis for most applications. Non-parametric tests such as the permutation test provide a powerful alternative by mimicking the distribution of the empirical data. For correlated data, the Fourier phase-randomization procedure~\cite{Theiler.1992} has been widely applied as it preserves temporal autocorrelation structure under stationarity assumptions. This standard method can be applied to the temporal dimension of our graph signals: 
$$
	\bbY = \bbX \bbF^{H} \bbPhi_\text{time} \bbF,
$$
where the diagonal of $\bbPhi_\text{time}$ contains random phase factors according to the windowing function $\Phi(\lambda_l)=\exp(j 2\pi \phi_l)$, with $\phi_l$ realizations\footnote{In practice, some additional constraints are added such as preservation of Hermitian symmetry.} of a random variable uniformly distributed in the interval $[0,1]$. From the surrogate signals, one can then compute a test statistic and establish its distribution under the null hypothesis by repeating the randomization procedure; i.e., the power spectrum density of the surrogate data is dictated by the empirical data. Note that in this setting, the spatial features of null realizations are identical to the ones of the actual data, while temporal non-stationary effects are destroyed.  

The phase randomization procedure can be generalized to the graph setting~\cite{pirondini1601} by considering the GFT. In particular, the graph signal can be decomposed on the GFT basis and then, the graph spectral coefficients can be randomized by flipping their signs.  
Assuming that the random sign flips are stored on the diagonal of $\bbPhi_\text{graph}$, we can formally write the procedure as:
\begin{align}\label{eqn_nulldata_graph}
    \bbY = \bbV \bbPhi_\text{graph} \bbV^\top \bbX.
\end{align}
This procedure generates surrogate graph signals in which the smoothness as measured on the graph is maintained, but in which the non-stationary spatial effects is destroyed. The temporal properties of null realizations are identical to those observed in the actual data.

%Once again, notice the strong similarity to \eqref{eqn_GFT_lowpass}.

%%%%%%%%%%%%%%%%%%%%%%%%%%%%%%%%%%%%%%%%%%%%%%%%%%%%%%%%%%%%%%%%%%%%%
%%%   S   U   B   S   E   C   T   I   O   N   %%%%%%%%%%%%%%%
%%%%%%%%%%%%%%%%%%%%%%%%%%%%%%%%%%%%%%%%%%%%%%%%%%%%%%%%%%%%%%%%%%%%%
%
\subsection{Wavelets and Slepians on the Graph}
\label{subsec:Slepians}
The wavelet transform is another fundamental tool of signal processing~\cite{Mallat.2009} providing localized, multiscale decompositions. Several designs have been proposed to generalize this concept to graphs, such as approaches in the vertex domain~\cite{Crovella.2003,Jansen.2009,Narang.2012}, based on diffusion processes~\cite{Coifman.2006,Talmon.2013}, or using the spectral domain~\cite{Hammond.2011,leonardi1302,behjat1601}. The latter design builds upon the GFT and has been applied for multiscale community mining~\cite{Tremblay.2014} or to investigate uncertainty principles~\cite{Agaskar.2013,pasdeloup2015,Tsitsvero.2016,Teke.2017}. 

Here, we detail a more recent design of a localized decomposition for graph signals that is based on a generalization of Slepian functions~\cite{vandeville1701} and that can deal with additional domain knowledge. %, which is particularly useful in the context of neuroimaging. 
Let us consider the problem of retrieving a signal $\bbx \in \mathbb{R}^{N}$ that is maximally concentrated within a subset of nodes from the graph at hand, while at the same time setting a maximal bandwidth on the solution. As the global concentration of a signal is given by $\bbx^\top \bbx$, we end up maximizing
\begin{align}\label{eqn_slepian_1}
    \mu = \frac{\tilde{\bbx}^\top \bar{\bbV}^\top \bbM \bar{\bbV} \tilde{\bbx}}{\tilde{\bbx}^\top \tilde{\bbx}},
\end{align}
where $\bbM$ is the diagonal \textit{selectivity matrix} with elements $M_{i,i} = 0$ or $1$ to respectively exclude, or include, a node into the sub-graph of interest, and $\bar{\bbV} \in \mathbb{R}^{N\times M}$ is a trimmed GFT matrix where only low-frequency basis vectors are kept. The interpretation here is that we aim at finding the linear combination of band-limited graph spectral coefficients enabling the best localization of the signal within the sub-graph. Note that the sub-graph is selected using prior information, and not optimized over.

If we define the \textit{concentration matrix} as $\bbC = \bar{\bbV}^\top \bbM \bar{\bbV}$, then the problem amounts to solving its eigendecomposition, and $\{\tilde{\bbs}_k\}, k = 0,1,\ldots,M-1$, are the weighting coefficients obtained as solutions. We assume that they are ordered in decreasing eigenvalue amplitude ($\mu_0 > \mu_1 > \ldots > \mu_{M-1}$), so that $\tilde{\bbs}_0$ is the optimal (maximally concentrated) solution. From the set of coefficients, the \textit{Slepian matrix} can then be retrieved as:
\begin{equation}
   \label{eqn_slepian_2}
    \bbS = \bar{\bbV} \tilde{\bbS},
\end{equation}
where $\bbS \in \mathbb{R}^{N\times M}$ and each column contains one of the Slepian vectors $\bbs_k$. Slepian vectors are not only orthonormal within the whole set of nodes ($\bbs_k^\top \bbs_l = \delta_{k-l}$), but also orthogonal over the chosen subset ($\bbs_k^\top \bbM \bbs_l = \mu_k \delta_{k-l}$).

Now, in order to make Slepian vectors more amenable to the application of GSP tools, let us consider an alternative optimization criterion in which the modified concentration matrix is given as $\mathbf{C_2} = \bar{\bbLambda}^{1/2} \bbC \bar{\bbLambda}^{1/2}$, with $\bar{\bbLambda} \in \mathbb{R}^{M\times M}$ the trimmed diagonal matrix of eigenvalues. The new quantity to optimize then reads:
\begin{align}\label{eqn_slepian_1b}
    \xi = \frac{\tilde{\bbx}^\top \bar{\bbLambda}^{1/2} \bar{\bbV}^\top \bbM \bar{\bbV} \bar{\bbLambda}^{1/2} \tilde{\bbx}}{\tilde{\bbx}^\top \tilde{\bbx}}.
\end{align}
The set of solution Slepian vectors are still orthonormal, but this time, they satisfy $\bbs_k^\top \bbM \bbs_l = \xi_k \delta_{k-l}$. Observe that, when using the Laplacian matrix as our graph shift operator, if all nodes are selected as the subset of interest ($\bbM = \bbI$) while enabling a full bandwidth ($\bar{\bbLambda} = \bbLambda$, $\bar{\bbV} = \bbV$), then we fall back on the classical Laplacian embedding case discussed in Section~\ref{subsec:GFT}, and as such, this modified criterion can be seen as a generalization of Laplacian embedding (i.e., a \textit{modified embedded distance} criterion) under user-defined bandwidth and selectivity constraints. 

\begin{figure}[t]
	\centerline{\includegraphics[width=0.49 \textwidth]{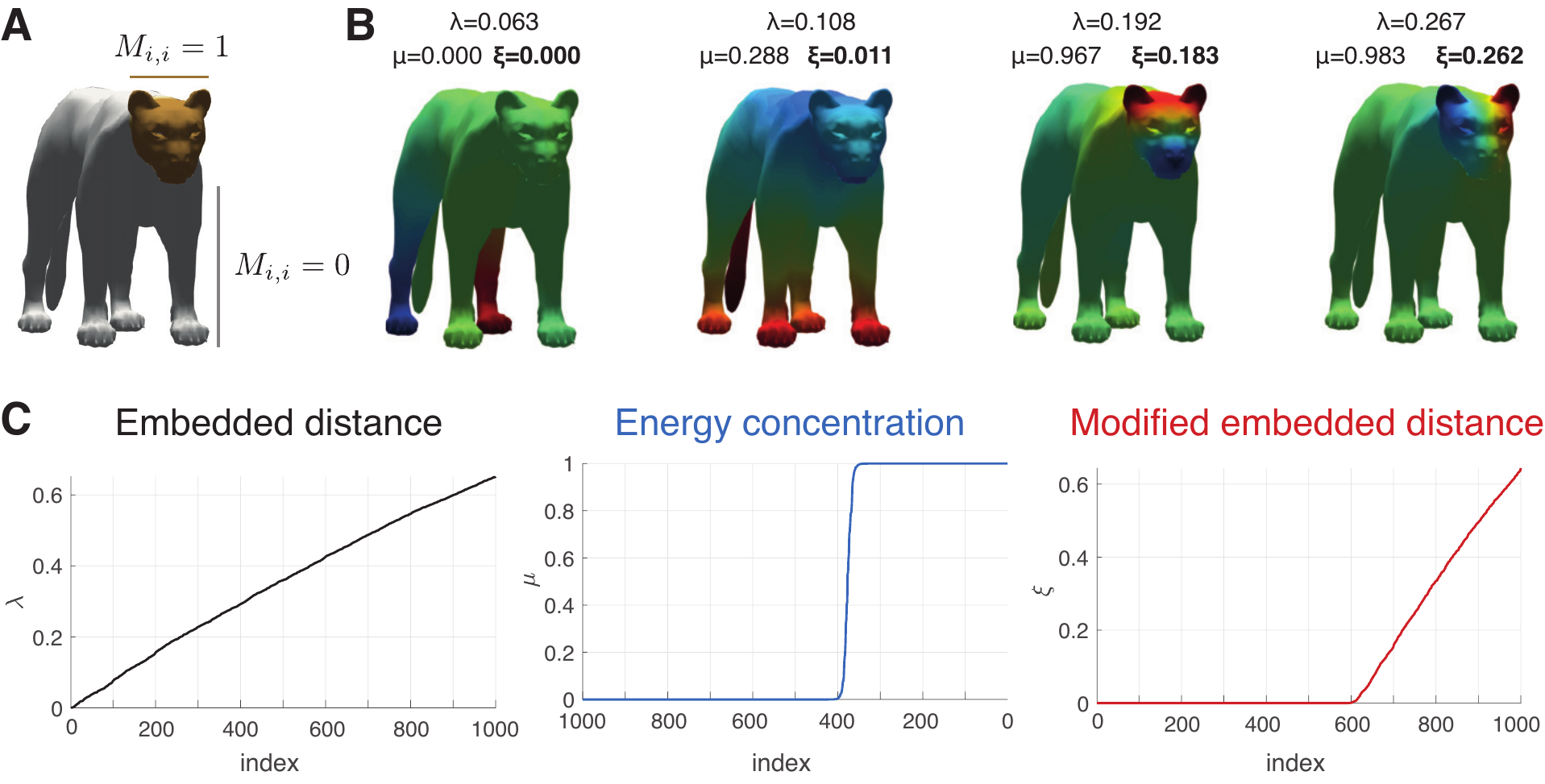}}
	\caption{\textbf{Illustration of Slepian vectors and their properties}. (\textbf{A}) Within the considered graph (a leopard mesh), the head is selected as the subset of nodes of interest. (\textbf{B}) Example Slepian vectors obtained from the modified embedded distance optimization criterion \eqref{eqn_slepian_1b}. In each case, alongside localized frequency ($\xi$), embedded distance ($\lambda$), and energy concentration ($\mu$) are also shown. (\textbf{C}) For a bandwidth $M = 1000$ and Laplacian embedding (left), energy concentration (middle) or modified embedded distance (right) optimizations, sorting of the obtained eigenvalues (respectively $\lambda$, $\mu$ or $\xi$).}
	\label{fig_slepians_intro}
\end{figure}

Analogously to the GFT setting, solution Slepian vectors of increasing eigenvalue $\xi_k$ can then be regarded as building blocks of increasing graph frequency, but within the chosen sub-graph, i.e., of increasing \textit{localized} frequency. The conceptual difference between both optimization schemes is illustrated in an example dataset of leopard mesh in Figure~\ref{fig_slepians_intro}, where the sub-graph is the head of the leopard as shown in Figure~\ref{fig_slepians_intro}A. Four of the Slepian vectors derived from \eqref{eqn_slepian_1b} are shown with their localized frequency $\xi$, their energy concentration $\mu$ computed from \eqref{eqn_slepian_1}, and their embedded distance $\lambda = \bbs^\top \bbL \bbs$. The leftmost example denotes a low frequency on the whole graph, with very weak signal within the selected sub-graph, and thus both low localized frequency and energy concentration. The second Slepian vector shows fairly uniform negative signal within the sub-graph, resulting in a quite large energy concentration, but a very low localized frequency. The last two examples reflect Slepian vectors that are both strongly concentrated (high $\mu$) and of high localized frequency (high $\xi$).

%%%%%%%%%%%%%%%%%%%%%%%%%%%%%%%%%%%%%%%%%%%%%%%%%%%%%%%%%%%%%%%%%%%%%
%%%   F   I   G   U   R   E   %%%%%%%%%%%%%%%
%%%%%%%%%%%%%%%%%%%%%%%%%%%%%%%%%%%%%%%%%%%%%%%%%%%%%%%%%%%%%%%%%%%%%
%

If Laplacian embedding is performed on the full graph (Figure~\ref{fig_slepians_intro}C, left plot), the resulting eigenvectors linearly span the graph frequency spectrum (black line). If the energy concentration criterion is used for generating Slepian vectors (middle plot), there is a well-defined transition point past which Slepian vectors become strongly concentrated within the selected subset of nodes. If the modified embedded distance criterion is used (right plot), then, past a point where Slepian vectors become concentrated within the subset (around 600 in this example), they also linearly span the localized graph frequency space.

As a result, it becomes possible to apply similar GSP tools as for the GFT, but for a decomposition that can be tailored in terms of localization by utilizing different subgraphs, and the choice of the bandwidth. In fact, the Slepian matrix can be seen as an alternative set of basis vectors, themselves obtained as a linear combination of Laplacian eigenvectors under the localized concentration constraint. For example, the temporal signal matrix $\bbX$ at hand can be projected on the Slepian building blocks as $\bbS^\top \bbX$, and if we define the diagonal matrix $\bbGamma_L$ as a localized low-pass filter by setting $[\bbGamma_L]_{i,i} = 1$ if $\xi_i < \hat{\xi_L}$ (low localized frequency) and $\mu_i > \epsilon$ (concentrated solution), or $0$ otherwise, the locally filtered output signal would be given by:
\begin{align}\label{eqn_slepian_3}
    \bbY_{\bbGamma_L} = \bbS \bbGamma_L \bbS^\top \bbX.
\end{align}

%%%%%%%%%%%%%%%%%%%%%%%%%%%%%%%%%%%%%%%%%%%%%%%%%%%%%%%%%%%%%%%%%%%%%
%%%   S   E   C   T   I   O   N   %%%%%%%%%%%%%%%
%%%%%%%%%%%%%%%%%%%%%%%%%%%%%%%%%%%%%%%%%%%%%%%%%%%%%%%%%%%%%%%%%%%%%
%
\section{A Brain GSP Case Study: Deciphering the Signatures of Attention Switching}
\label{sec:applications}

\begin{figure}[t]
	\centerline{\includegraphics[width=0.49 \textwidth]{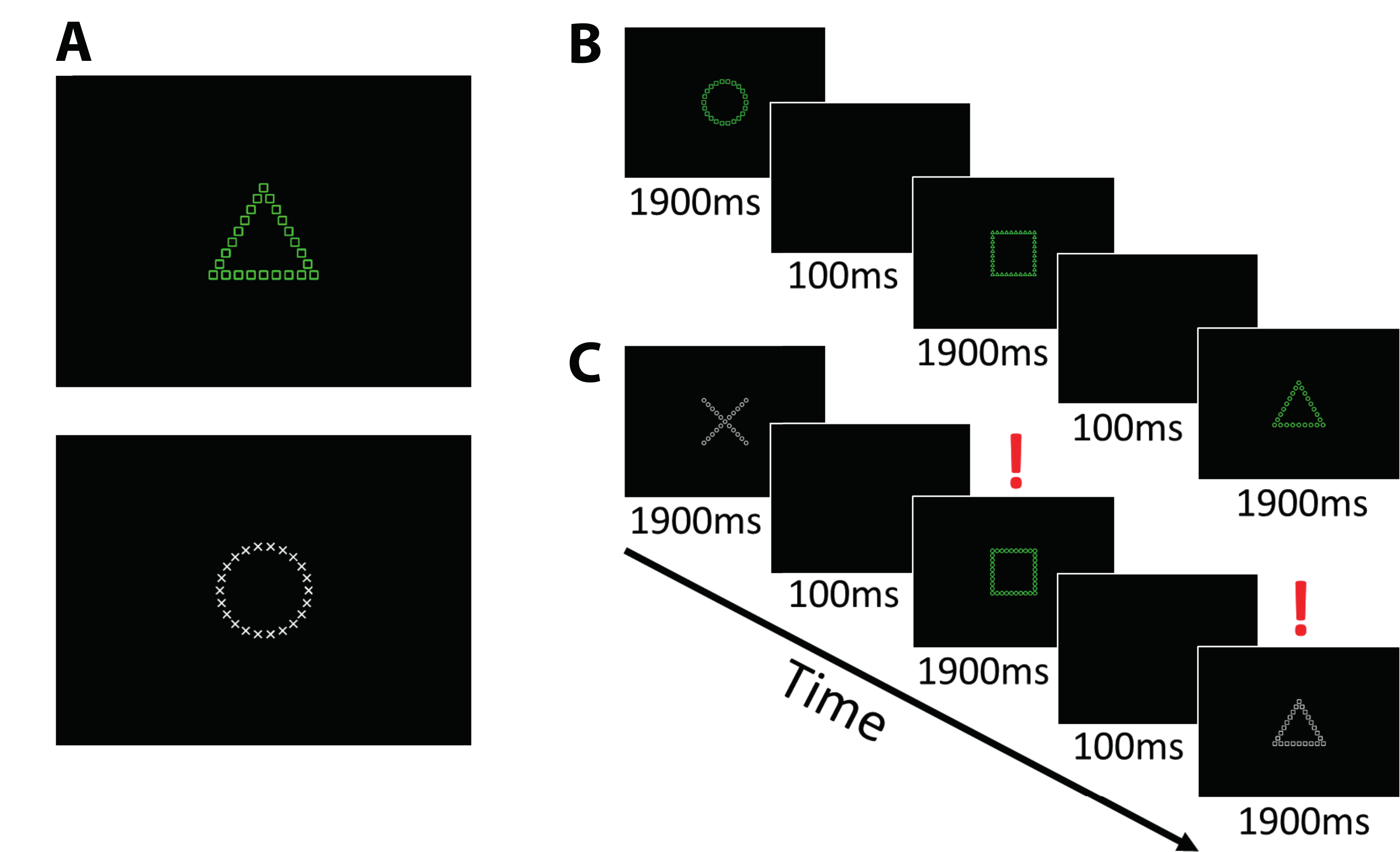}}
	\caption{\textbf{Cognitive task requiring perceptual switching}. (\textbf{A}) Example stimuli based on Navon local-global features. Subjects were trained to respond to the larger (or ``global'') shapes if the stimulus was green and to the smaller (or ``local") shapes if it was white. (\textbf{B}) An example of the non-switching condition for responses. Subjects viewed a sequence of images and were instructed to respond as quickly and as accurately as possible. (\textbf{C}) An example of the switching condition between stimuli requiring global and local responses. Here, trials with a red exclamation mark are switches from the previous stimulus. [Reproduced with permission from \cite{medaglia2016}.]\vspace{-2mm}} 
	\label{fig_attention_switching}
\end{figure}

\begin{figure*}[t]
	\centerline{\includegraphics[width=1 \textwidth, trim=0cm 0.0cm 0cm 0.0cm, clip=true]{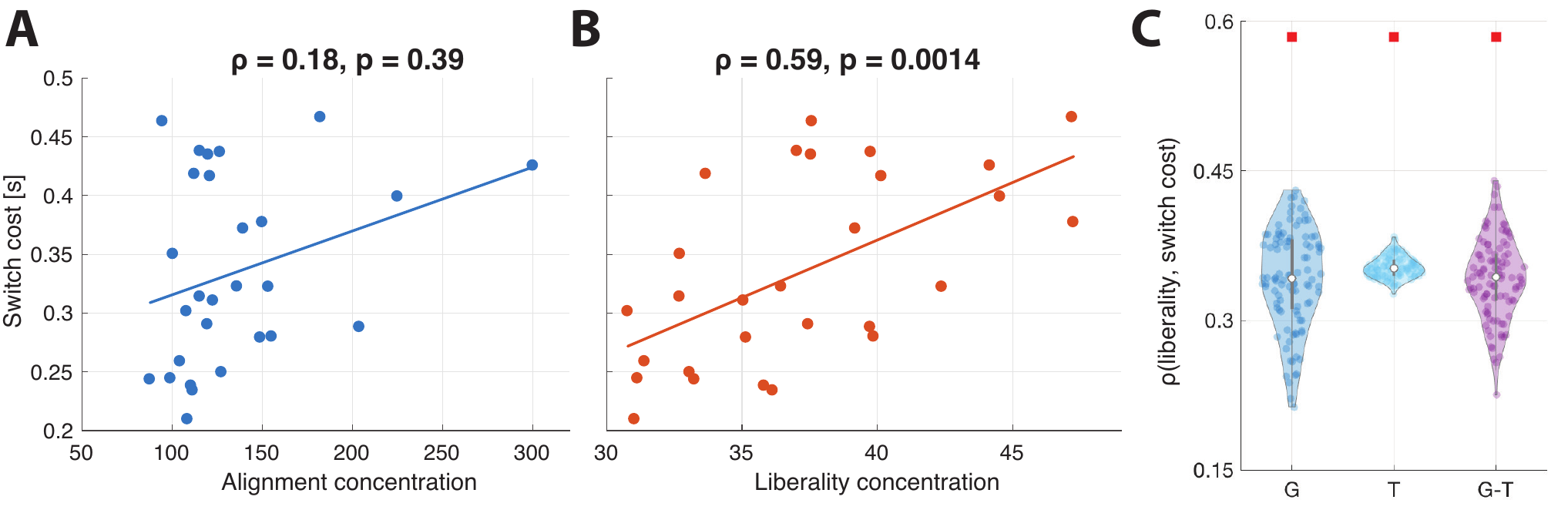}}
	\caption{\textbf{Switch cost correlates with the concentration in liberal signal}. (\textbf{A}) Switch cost does not significantly relate to the concentration of the low-frequency functional signal component (alignment). (\textbf{B}) A lower concentration of graph high-frequency components is associated with a lower switch cost, that is, with faster attention switching. (\textbf{C}) The correlation between switch cost and liberal signal concentration is much stronger in the actual data than in null realizations, irrespective of whether the statistical randomization is performed in the graph domain (denoted as 'G' in the figure), in the temporal domain (denoted as 'T' in the figure), or jointly performed in both (denoted as 'G-T' in the figure). Blue, cyan and purple data points denote the correlation coefficients obtained from surrogate signals under the three null models, while the red rectangle indicates the real correlation coefficient ($\rho = 0.59$). $\rho$, partial Pearson's correlation coefficient; $p$, p-value. [A and B are adapted with permission from \cite{medaglia2016}].} 
	\label{fig_switching_results}
\end{figure*}

We now discuss how the aforementioned GSP methods can be applied in the context of functional brain imaging. Figure \ref{fig_attention_switching} is reproduced from \cite{medaglia2016}; Figures \ref{fig_switching_results}A and B are adapted from \cite{medaglia2016}. To do so, we focus on the data whose acquisition was described in Section~\ref{sec_graphs_and_signals}, Callouts. For each volunteer, fMRI recordings were obtained when performing a Navon switching task, where local-global perception is assessed using classical Navon figures \cite{navon1977forest}. Local-global stimuli were comprised of four shapes -- a circle, cross, triangle, or square -- that were used to build the global and local aspects of the cues (see Figure~\ref{fig_attention_switching}A for examples).

A response (button press) to the local shape was expected from the participants in the case of white stimuli, and to the global shape for green ones. Two different block types were considered in the experiment: in the first one (Figure~\ref{fig_attention_switching}B), the color of the presented stimuli was always the same, and the subjects thus responded consistently to the global or to the local shapes. In the second block type (Figure~\ref{fig_attention_switching}C), random color switches were included, so that slower responses were expected. The difference in response time between the two block types, which we refer to as \textit{switch cost}, quantifies the behavioral ability of the subjects.

%%%%%%%%%%%%%%%%%%%%%%%%%%%%%%%%%%%%%%%%%%%%%%%%%%%%%%%%%%%%%%%%%%%%%
%%%   F   I   G   U   R   E   %%%%%%%%%%%%%%%
%%%%%%%%%%%%%%%%%%%%%%%%%%%%%%%%%%%%%%%%%%%%%%%%%%%%%%%%%%%%%%%%%%%%%
%

To study the association between brain signal and attention switching, we decomposed the functional brain response into two separate components: one representing \textit{alignment} with structural connectivity (i.e., the regions that activate together are also physically wired), and one describing \textit{liberality} (i.e., the areas that exhibit high signal variability with respect to the underlying graph structure). To do so, we performed graph signal filtering (Section~\ref{subsec:graph_filtering}) with two different filtering matrices: (1) $\bbPsi_\text{Al}$, so that $\bbY_{\bbPsi_\text{Al}} = \bbV \bbPsi_\text{Al} \bbV^\top \bbX$ is the transformed (low-pass filtered) functional data in which only the $10$ lowest frequency modes are expressed at each time point; and (2) $\bbPsi_\text{Lib}$, for which $\bbY_{\bbPsi_\text{Lib}}$ only represents the temporal expression of the $10$ largest frequency modes (high-pass filtering). At a given time point, the filtered functional signal varies in sign across brain regions. Thus, to derive a subject-specific scalar quantifying alignment or liberality, we considered the norms of those signals as measures of concentration, which were eventually averaged across all temporal samples of a given subject. We used the $\ell_2$ norm because it provides an interpretation of energy for each graph frequency component; other reasonable choices of norm, including the $\ell_1$ norm, yield similar results. Also, presented results are obtained using the adjacency matrix as the graph shift operator, but similar findings were recovered using the Laplacian matrix instead (see Callout 3).

%%%%%%%%%%%%%%%%%%%%%%%%%%%%%%%%%%%%%%%%%%%%%%%%%%%%%%%%%%%%%%%%%%%%%
%%%   M   A   I   N      M   A   T   T   E   R   %%%%%%%%%%%%%%%
%%%%%%%%%%%%%%%%%%%%%%%%%%%%%%%%%%%%%%%%%%%%%%%%%%%%%%%%%%%%%%%%%%%%%
To relate signal alignment and liberality to cognitive performance of the participants, we computed partial Pearson's correlation between our concentration measures and switch cost (median additional response time during switching task blocks compared to non-switching task blocks). Age and motion were included as covariates to remove their impact from the results. Regarding alignment, there was no significant association ($p > 0.35$; Figure~\ref{fig_switching_results}A). In other words, the extent with which functional brain activity was in line with the underlying brain structural connectivity did not relate to cognitive abilities in the assessed task. However, we observed a significant positive correlation between liberal signal concentration and switch cost ($\rho = 0.59$, $p < 0.0015$; see Figure~\ref{fig_switching_results}B). Thus, the subjects exhibiting most liberality in their functional signals were also the ones for whom the attention switching task was the hardest. We verified that the high-frequency modes involved in those computations were not solely localized to a restricted set of nodes by evaluating the distribution of the average decomposed signal across all brain regions. When averaged across all time points and subjects, 27 brain regions had their decomposed signals higher than 1.5 times the mean of the distribution (approximately 3), confirming that a wide area of the brain was spanned by high-frequency modes. From these results, one can see that a GSP framework %enables to isolate the functional components that are responsible for faster attention switching.
may provide a way to disentangle brain signals that exhibit different levels of association with attention switching.

%%%%%%%%%%%%%%%%%%%%%%%%%%%%%%%%%%%%%%%%%%%%%%%%%%%%%%%%%%%%%%%%%%%%%
%%%   F   I   G   U   R   E   %%%%%%%%%%%%%%%
%%%%%%%%%%%%%%%%%%%%%%%%%%%%%%%%%%%%%%%%%%%%%%%%%%%%%%%%%%%%%%%%%%%%%
%

To more thoroughly examine the significance of the association between liberal signals and switch cost, we performed a null permutation test by generating graph surrogate signals as described in Section~\ref{subsec:SpectralSurrogate}. Specifically, we generated $100$ graph surrogate signals by randomly flipping the signs stored on the diagonal of $\bbPhi_\text{graph}$, as in \eqref{eqn_nulldata_graph}. Then, we evaluated the association between the null surrogate signals and switch cost. As seen in Figure~\ref{fig_switching_results}C (case `G'), the actual correlation coefficient between liberal signal concentration and switch cost (denoted by the red rectangle) is significantly larger than when computed on any of the null graph surrogate signals. We also performed the same process using phase randomization in the time domain to generate surrogate signals (see Figure~\ref{fig_switching_results}C, case `T'), which preserves the temporal stationarity assumption, and combining phase randomization in the time domain and randomly flipping the signs of graph spectral coefficients (Figure~\ref{fig_switching_results}C, case`G-T'). Again, the actual correlation coefficient between liberal signal concentration and switch cost was significantly larger than for any of the null realizations. 

To confirm that the graph frequency decomposition framework is insensitive to the level of resolution used in the considered parcellation, we examined the data recorded during the same experiment, on the same subjects, but at a higher resolution ($N = 262$ different brain regions). In other words, we considered the same experiment, but defined the network differently by having each node of $\ccalV$ consisting of a smaller volume of the brain. We followed the same graph frequency decomposition, using the adjacency matrix as graph shift operator, on this finer graph. We observed that the results still held, as switch cost did not significantly relate to the concentration of the low-frequency signal component ($\rho = 0.3408, p = 0.0759$), whereas a lower concentration of the high-frequency component was associated with faster attention switching ($\rho = 0.4232, p = 0.0249$). Here and above, the results were also robust to the number of largest/smallest frequency components used in the decomposition.

In sum, in this section we reviewed a recent study \cite{medaglia2016} demonstrating that individuals whose most liberal fMRI signals were more aligned with white matter architecture could switch attention faster. In other words, relative alignment with anatomy is associated with greater cognitive flexibility. This observation complements prior studies of executive function that have focused on node-level, edge-level, and module-level features of brain networks \cite{braun2015dynamic, leunissen2014subcortical}. The importance of this finding illustrates the usefulness of GSP tools in extracting relevant cognitive features.
%We found that the liberal signals sampled from any block type were correlated with response times and switch cost of the trials that occurred during switching blocks. Liberal signals were not correlated with performance on the non-switching blocks, suggesting that these signals are specifically related to cognitive control demands introduced during the switching condition. The fact that this relationship can be found between switching performance and the signals calculated in the fixation, non-switching, and switching blocks suggests that liberal signal organization is a trait-like variable existing in both periods of switching tasks and rest. This may indicate that some brains are at a natural advantage to overcome switching demands, and that behaviorally-relevant connectomic signal embedding in anatomy can be detected across the task conditions.

%%%%%%%%%%%%%%%%%%%%%%%%%%%%%%%%%%%%%%%%%%%%%%%%%%%%%%%%%%%%%%%%%%%%%
%%%   M   A   I   N      M   A   T   T   E   R   %%%%%%%%%%%%%%%
%%%%%%%%%%%%%%%%%%%%%%%%%%%%%%%%%%%%%%%%%%%%%%%%%%%%%%%%%%%%%%%%%%%%%
%

\begin{figure*} [t]
\centerline{\includegraphics[width=1 \textwidth]{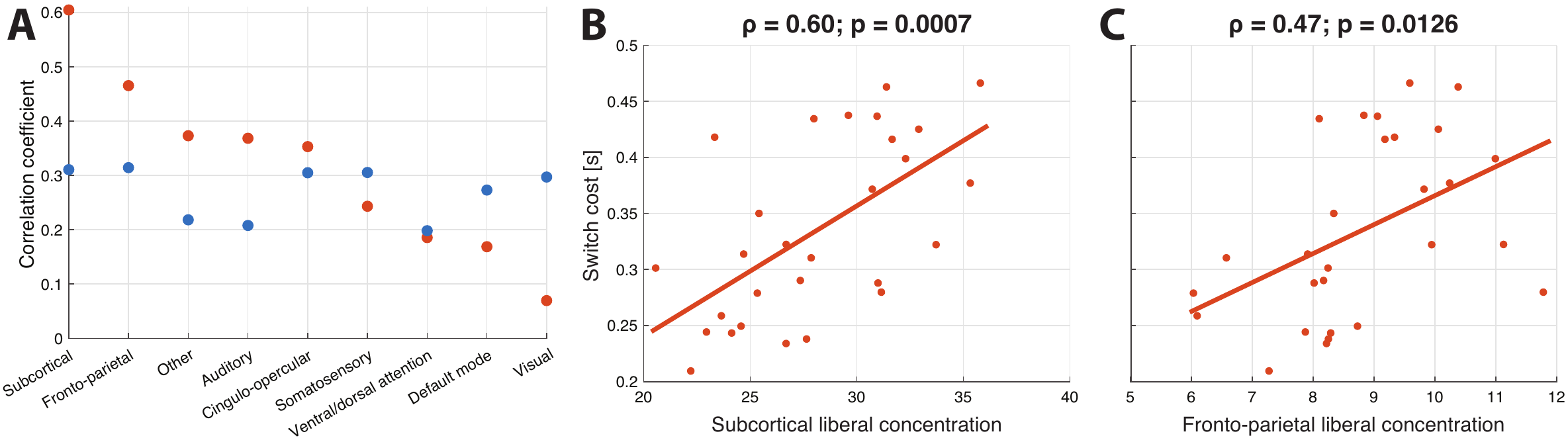}} 
	\caption{\textbf{Pinpointing the brain systems involved in attention switching}. (\textbf{A}) Separate partial correlation assessments between switch cost and alignment (blue) or liberality (red) signal concentration on the brain areas belonging to different functional systems, using age and motion as covariates. Systems are ordered in decreasing liberality correlation coefficient order. Liberality concentrations of subcortical and fronto-parietal systems exhibit the highest and most significant contributions to the association with switch cost. Liberality concentrations of other systems and alignment concentrations of any system exhibited no significant association ($p > 0.05$). (\textbf{B}) A lower concentration of graph high-frequency components in the subcortical system is associated with faster attention switching. (\textbf{C}) A lower concentration of graph high-frequency components in the fronto-parietal system is associated with faster attention switching. \vspace{-2mm}}
	\label{fig_correlation_by_region}
\end{figure*}

Up to this point, we have been dealing with a graph frequency decomposition considered at the level of the whole brain. However, GSP tools also allow us to independently evaluate separate nodes, or sets of nodes, from the graph at hand. In the present case, this flexibility permits a more in-depth study of which brain regions are specifically responsible for the observed association between liberality and switch cost. For this purpose, we considered 9 different, previously defined functional brain systems \cite{Gu2015}, each of which included a distinct set of regions. We assessed, separately for each system, the correlation between switch cost and alignment or liberality. In the former case (alignment), there was no significant association, whereas in the latter (liberality), the relationship seen in Figure~\ref{fig_switching_results}B could be narrowed down to two significant contributors: the subcortical and the fronto-parietal systems (Figure~\ref{fig_correlation_by_region}). Those results highlight the ability of GSP tools to not only decompose signals in the graph \textit{frequency domain}, but also in the graph \textit{spatial domain} (examining different nodes in the graph). Combining those two analytical axes enables us to gather deeper insights into functional brain activity and its relation to cognition.

%%%%%%%%%%%%%%%%%%%%%%%%%%%%%%%%%%%%%%%%%%%%%%%%%%%%%%%%%%%%%%%%%%%%%
%%%   F   I   G   U   R   E   %%%%%%%%%%%%%%%
%%%%%%%%%%%%%%%%%%%%%%%%%%%%%%%%%%%%%%%%%%%%%%%%%%%%%%%%%%%%%%%%%%%%%
%

%%%%%%%%%%%%%%%%%%%%%%%%%%%%%%%%%%%%%%%%%%%%%%%%%%%%%%%%%%%%%%%%%%%%%
%%%   S   U   B   S   E   C   T   I   O   N   %%%%%%%%%%%%%%%
%%%%%%%%%%%%%%%%%%%%%%%%%%%%%%%%%%%%%%%%%%%%%%%%%%%%%%%%%%%%%%%%%%%%%
%
\section{Perspectives for Brain GSP: Studying Functional Dynamics}
\label{sec:dynamics}

\subsection{Resolving excursions in alignment or liberality regimes}
\label{subsec:excursions}

%%%%%%%%%%%%%%%%%%%%%%%%%%%%%%%%%%%%%%%%%%%%%%%%%%%%%%%%%%%%%%%%%%%%%
%%%   F   I   G   U   R   E   %%%%%%%%%%%%%%%
%%%%%%%%%%%%%%%%%%%%%%%%%%%%%%%%%%%%%%%%%%%%%%%%%%%%%%%%%%%%%%%%%%%%%
%
\begin{figure*}
  \centering
  \includegraphics[width=0.98\textwidth]{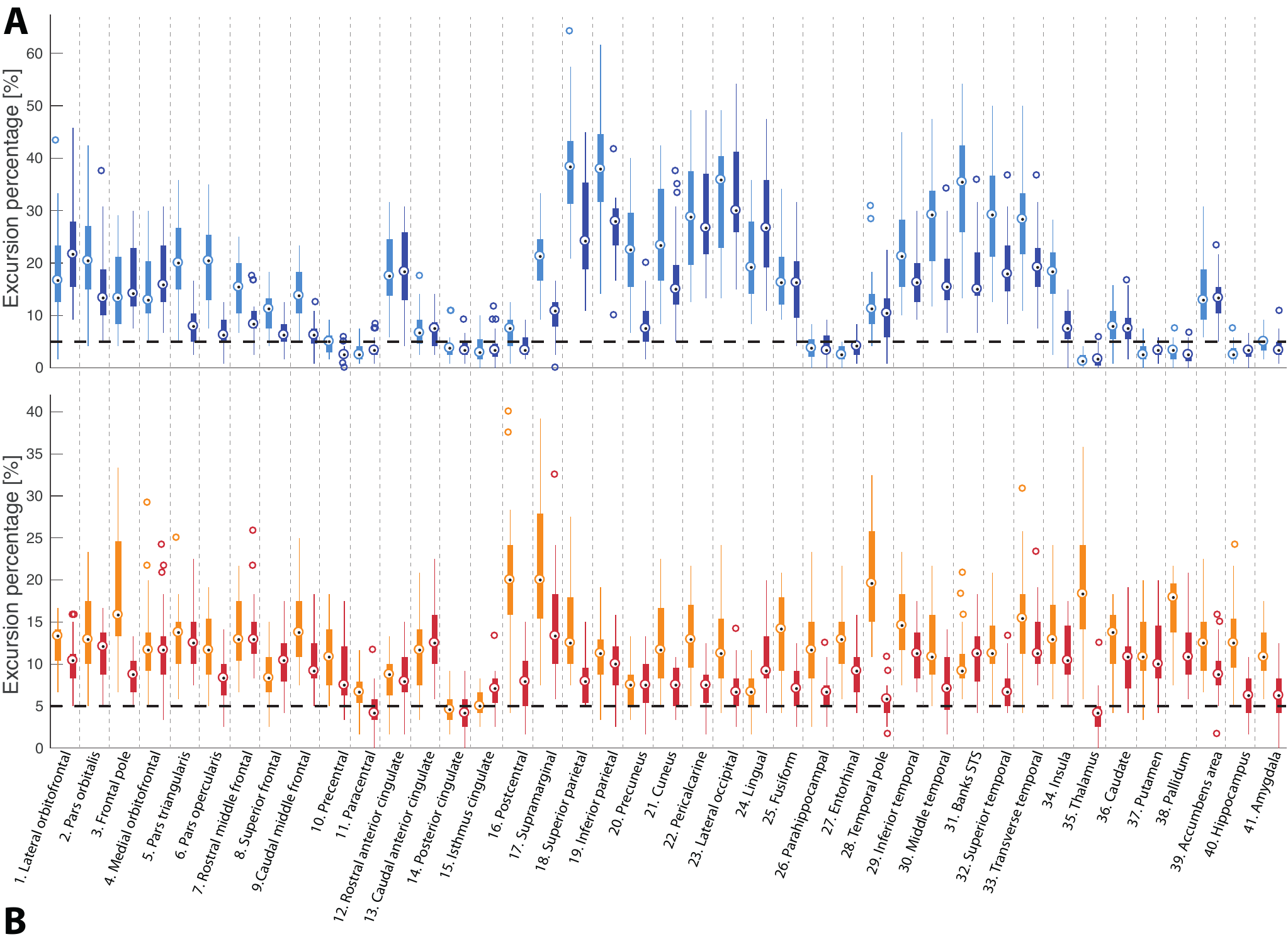}\\[1.5ex]
  \fcolorbox{black}{blue!10}{
     \includegraphics[width=0.16\textwidth]{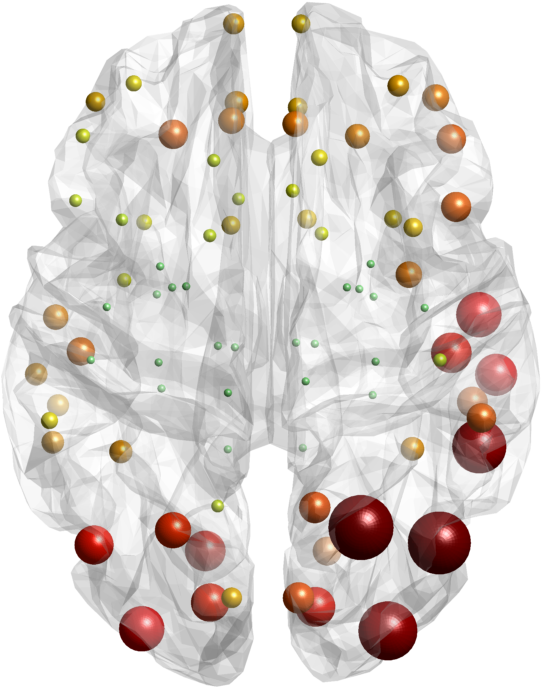}
     \includegraphics[width=0.27\textwidth]{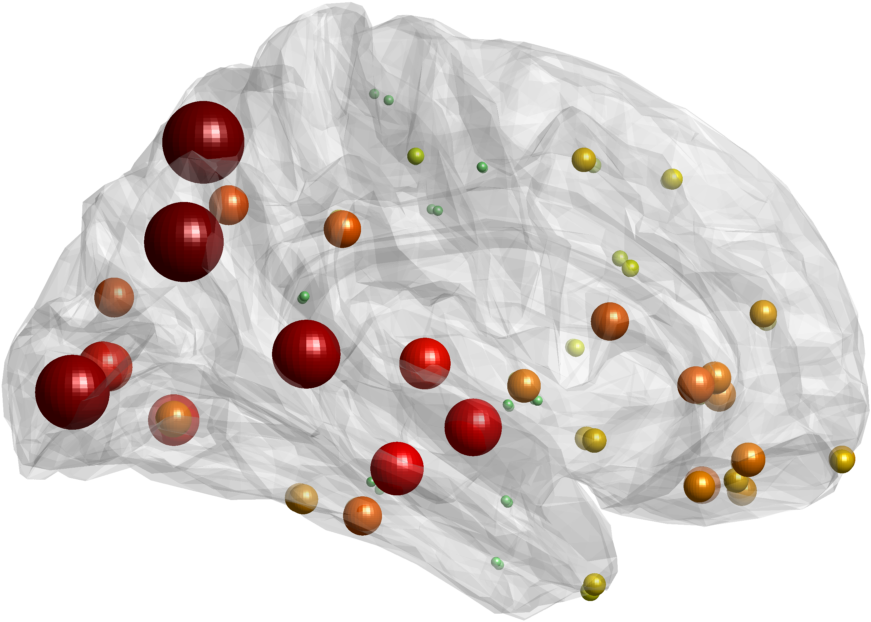}
  }
  \fcolorbox{black}{red!10}{
     \includegraphics[width=0.16\textwidth]{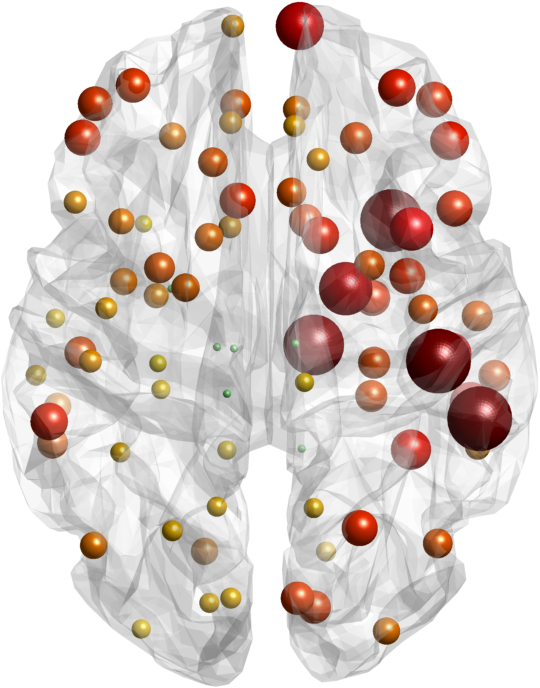}
     \includegraphics[width=0.27\textwidth]{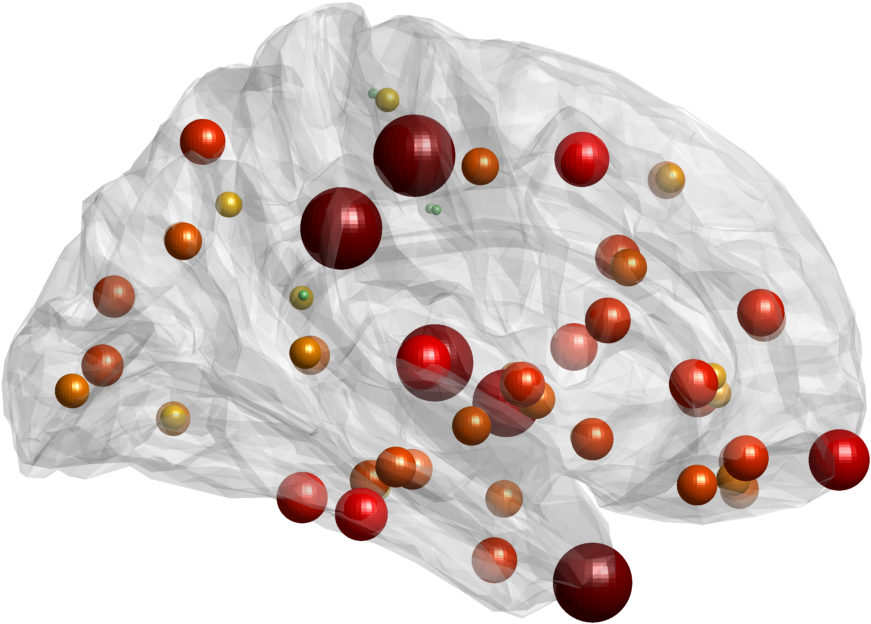}
  }
  \caption{\label{fig:RS_regions} \textbf{Significant excursions of aligned and liberal functional signals across regions}. (\textbf{A}) For all $82$ nodes, percentage of significant excursions for alignment (top panel, light and dark blue box plots) or liberality (bottom panel, red and orange box plots) across subjects. The horizontal dashed line denotes chance level ($\alpha = 5\%$), and light gray vertical dashed lines separate the box plots from different regions. Light colors denote regions from the left side of the brain, and dark colors from the right side. (\textbf{B}) For alignment (left box) and liberality (right box), horizontal and sagittal brain views depicting excursion occurrence across brain nodes. A larger amount of significant excursions is denoted by a bigger and redder sphere. Left on the brain slices stands for the right side of the brain.}
\end{figure*}

We now illustrate, on the same data as above, how GSP tools can be applied to provide insights into the dynamics of functional brain activity. For every subject, we generated $1000$ null signal matrices using the strategy outlined in \eqref{eqn_nulldata_graph} (graph domain randomization). We combined this operator ($\bbPhi_\text{graph}$) with the alignment/liberality filtering operations, to generate null data for the aligned and liberal signal components. Formally, we thus computed a null realization as $\bbY = \bbV \bbPhi_{\text{graph}} \bbPsi_\text{Al} \bbV^\top \bbX$ or $\bbY = \bbV \bbPhi_{\text{graph}} \bbPsi_\text{Lib} \bbV^\top \bbX$, respectively. At an $\alpha$-level of $5\%$, we then used the generated null data to threshold the filtered signals, in order to locate significant signal \textit{excursions} -- particular moments in time when entering a regime of strong alignment, or liberality, with the underlying brain structure. In doing so, we considered absolute graph signals. Presented results are obtained using the adjacency matrix as graph shift operator, but similar findings were recovered using the Laplacian matrix (see Callout 3).

Figure~\ref{fig:RS_regions}A highlights the percentage of time points showing significant excursions for the aligned (light blue and dark blue box plots) and liberal (red and orange box plots) signal components across brain regions. An excursion percentage value of $5\%$ (horizontal dashed line) denotes chance level. Such a case was, for instance, observed for the paracentral and posterior cingulate areas (nodes 11 and 14), both in terms of aligned and liberal signal contributions. As null data realizations were generated in the graph domain, this observation means that those nodes did not show signal fluctuations going beyond what could be accounted for by the underlying spatial smoothness of the brain's structural graph.

Most brain regions did display very significant excursion percentages: considering alignment, occipital (nodes 21-25), parietal (nodes 18 and 19) and temporal (nodes 29-33) regions were the strongest contributors, while for liberality, key areas were located in temporal (nodes 29-33), subcortical (nodes 34, 36-39) or frontal (nodes 1-9) regions. Figure~\ref{fig:RS_regions}B displays the anatomical location of the main contributing regions. Qualitatively similar findings were also obtained when resorting to a finer parcellation of the brain ($N=262$ regions; see Supplementary Figure 1). The observation that the majority of brain nodes show frequent moments of strong alignment or liberality with respect to brain structure is consistent with current knowledge on spontaneous brain dynamics, since an alternation between time points with and without global similarity to the structural scaffold has previously been documented from second-order connectivity analyses \cite{Betzel2016,Liegeois2016}. A GSP approach can also reveal these subtle relationships, with the added advantage of conserving a frame-wise temporal resolution.

To better grasp the signal features at the root of alignment or liberality excursions, we compared the outcomes obtained using the graph surrogate method to the ones generated with the more classical Fourier phase-randomization procedure to generate null data, or to the outcomes resulting from the combination of those two surrogate approaches (see Supplementary Figure 2). Excursions in terms of liberality with respect to brain structure were not resolved anymore under those two other null models, for which null realizations conserve similar stationary temporal properties. This implies that the liberal signal component can be explained by stationary temporal features. On the other hand, alignment excursions remained, in particular when including graph domain randomization. Thus, the aligned signal component relates to spatial features that cannot be explained by stationary smoothness alone.

\subsection{Combining graph excursions with Fourier analysis}
\label{subsec:further}

\begin{figure*}
  \centering
  \includegraphics[width=\textwidth]{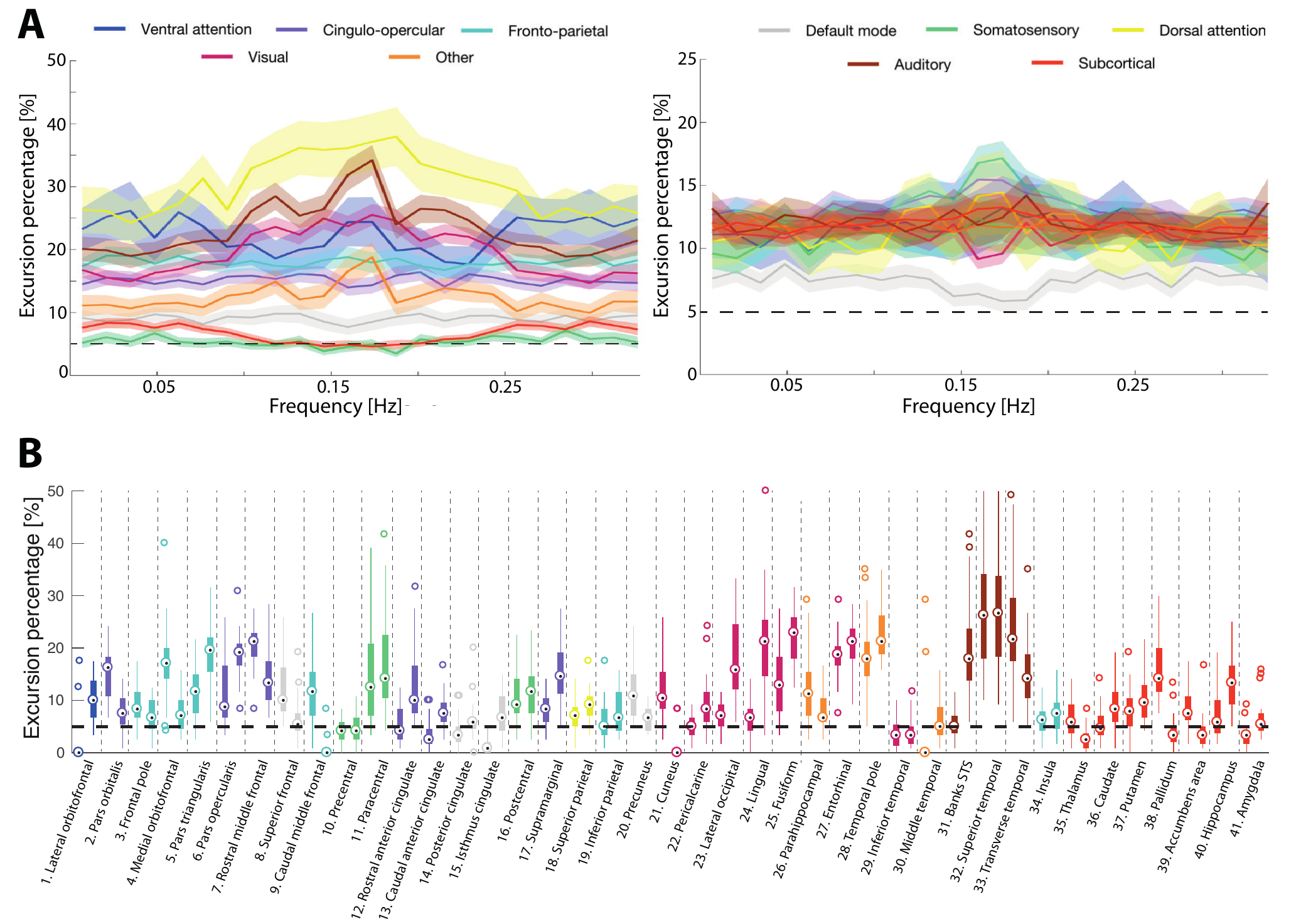}
  \caption{\label{fig:RS_systems} \textbf{Further disentangling functional brain signals by more elaborate GSP building blocks}. (\textbf{A}) Percentage of significant excursions for key functional brain systems across temporal frequency sub-bands in the case of the aligned (left graph) or liberal (right graph) signal contributions. Two-tailed $95\%$ confidence intervals are displayed for each curve, and the horizontal dashed line represents the excursion chance level ($\alpha = 5\%$). (\textbf{B}) As a quantification of local alignment, percentage of significant excursions for all brain nodes when applying the graph Slepian design with bandwidth $M=80$. Color coding reflects the functional system to which a region belongs, and for a given region, the left box plot stands for the left side of the brain.}
\end{figure*}

Other ingredients from the GSP pallet can be appended to the pipeline we have introduced, in order to further expand our understanding of brain activity. For example, to examine whether alignment and liberality would change along frequency, referring this time to the \textit{temporal frequency} of the signal, we simply combined our null and alignment/liberality operators with the classical Fourier decomposition highlighted in Section~\ref{subsec:GFT}, and computed the percentage of significant excursions for all the functional brain systems introduced in \cite{Gu2015} (Figure~\ref{fig:RS_systems}A). For alignment (left graph), different systems were observed to vary in terms of excursion occurrence, with dorsal attention and auditory areas as primary contributors while subcortical and somatosensory regions stood at around chance level. Interestingly, in a few cases, alignment with the structural brain scaffold appeared to be maximized at particular frequencies: for instance, the dorsal attention, ventral attention and auditory systems showed more frequent excursions in the $0.15-0.2$Hz range.

Regarding liberality (right graph), almost all systems showed similar excursion percentages, with the exception of the default mode network (gray line), whose regions appeared to more rarely diverge from the activation patterns expected from structural connectivity. In addition, excursions further decreased close to chance level in the $0.15-0.2$Hz range, while at the same time, positive peaks could be seen, amongst others, for the fronto-parietal and cingulo-opercular systems. This antagonistic relationship between those functional brain systems could be the reflection of a hallmark feature of brain activity: the anti-correlation between the default mode (also known as \textit{task-negative}) and so called \textit{task-positive} networks \cite{Fox2005}. The GSP approach enables, a more accurate characterization of these networks in terms of both temporal and graph frequencies.

\subsection{Probing excursions within a sub-graph with Slepians}
\label{subsec:further2}

Finally, another way to dig deeper into the functional signals is to consider them at a \textit{local scale}, rather than at the whole-brain level. For this purpose, we computed a basis of Slepian vectors through the process detailed in Section~\ref{subsec:Slepians} (using the modified embedded distance optimization criterion). We started from the eigendecomposition of the Laplacian matrix, and iteratively focused the analysis on a subset of nodes being part of only one given functional brain system. Every time, we derived $M=80$ Slepian vectors, and used the $10$ lowest localized frequency (i.e., with lowest $\xi_i$), concentrated (i.e., satisfying $\mu_i > \epsilon$) elements of this new basis to extract the part of the functional signals aligned with local structural brain features, generate null data, and quantify significant excursions.

As can be seen in Figure~\ref{fig:RS_systems}B, some nodes stand out as undergoing particularly frequent excursions in terms of local alignment to brain structure. This is for example seen for regions from the visual (nodes 23-25) and auditory (nodes 31-33) systems, reflecting the presence of moments when there is strong alignment of the functional signals with the underlying structure \textit{at the local scale of the considered system}, which is encoded in the Slepian basis. We note that the same nodes already showed high excursion percentages in Figure~\ref{fig:RS_regions}A, where alignment was assessed at the global (not local) level, and thus, what was captured by this less focused analysis may have largely involved local alignment with structure. Conversely, there are also many cases in which regions exhibited frequent global alignment with the structural scaffold, without displaying it at the local scale (for example, nodes 18-19). In such cases, global alignment to structure instead reflects cross-network interactions. Overall, surrogate analyses are conducted from three aspects in the preceding subsections (vanilla as in Section \ref{subsec:excursions}, combined with Fourier analysis as in Section \ref{subsec:further}, and combined with Slepians sub-graph as in Section \ref{subsec:further2}). Some consistent observations inherited from the surrogate analysis itself are found across the subsections, while some different results reflect the different perspectives and features of the respective approach.

\begin{figure*}
    \begin{framed}
    \begin{center}\sc Callout 3: Impact of the Graph Shift Operator.\end{center} \small
    Multiple graph shift operators could be used to decompose graph signals. Most of the material presented in this work uses the adjacency matrix as graph shift operator, but results remain very similar if the Laplacian matrix is used instead. More specifically, we reevaluated the association with switch cost illustrated in Figure~\ref{fig_switching_results}, and the set of brain regions most frequently undergoing alignment or liberality excursions as displayed in Figure~\ref{fig:RS_regions}B, using the Laplacian matrix as graph shift operator. Figure~\ref{fig_correlation_laplacian}, presented below, illustrates the similarity in the obtained results. There exist other types of graph shift operators, e.g. the normalized Laplacian, for which results can also be expected to remain relatively similar. 
    \end{framed}
    
    \centerline{\includegraphics[width=1\textwidth]{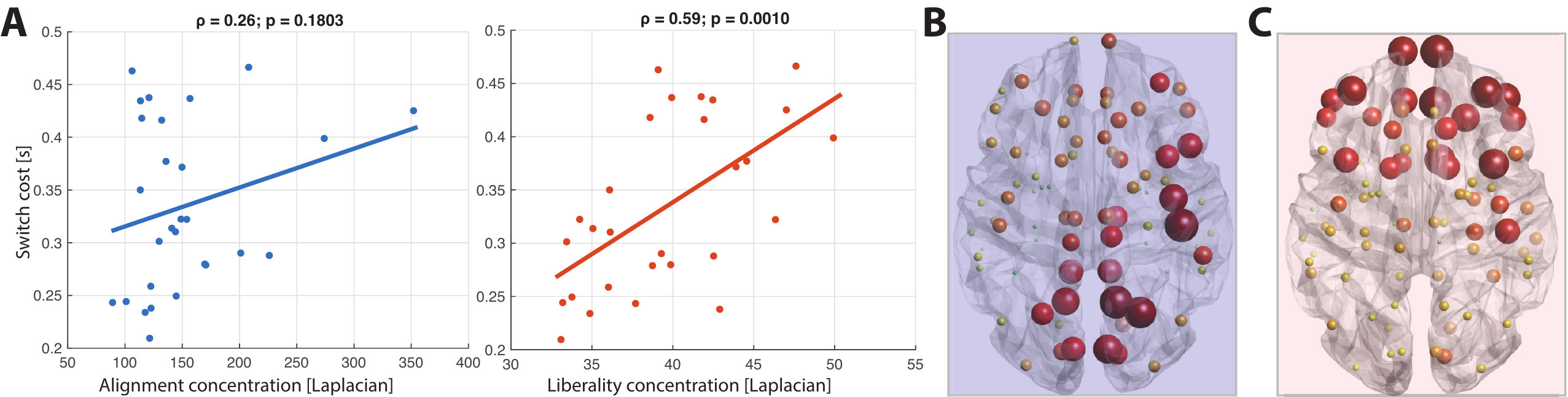}}
    \caption{\textbf{Results carry over to alternative graph shift operator -- graph Laplacian} (\textbf{A}) Switch cost correlates with the concentration in liberal signal and not aligned signal using Laplacian as a shift operator; results are similar as in Figure \ref{fig_switching_results}A and B. Horizontal brain views depicting excursion occurrence across brain nodes with Laplacian for alignment (\textbf{B}) and liberality (\textbf{C}); results are similar as in Figure \ref{fig:RS_regions}B.}
     \label{fig_correlation_laplacian}
\end{figure*}

%%%%%%%%%%%%%%%%%%%%%%%%%%%%%%%%%%%%%%%%%%%%%%%%%%%%%%%%%%%%%%%%%%%%%
%%%   S   E   C   T   I   O   N   %%%%%%%%%%%%%%%
%%%%%%%%%%%%%%%%%%%%%%%%%%%%%%%%%%%%%%%%%%%%%%%%%%%%%%%%%%%%%%%%%%%%%
%

\section{Conclusion \& Perspectives}
\label{sec:perspectives}
The GSP framework enables the analysis of brain activity on top of the structural brain graph. In particular, we have analyzed anatomically aligned or liberal organization of brain activity, and in the context of an attention switching task, we have reviewed a recent study \cite{medaglia2016} that signals aligned with anatomical connectivity are the most variable over time in cingulo-opercular and fronto-parietal systems; see \cite{medaglia2016} for a more detailed discussion. In addition, we used surrogate signals
to generate graph null models to suggest that the significance of the
results cannot be explained by random permutations. These results reinforce similar findings that were based on functional graphs~\cite{huang2016}, where we used the same approach to decompose fMRI signals based on dynamic functional connectivity and observed that different graph frequency components exhibited different importance depending on whether subjects were familiar or unfamiliar with the underlying task. Unlike conventional signal processing where low frequency is typically considered as information and high frequency considered as noise, we notice that in applying graph signal processing, both graph low and high frequencies may contain highly valuable information.

In addition to our review of attention switching, we have also introduced possible avenues for the use of GSP tools in uncovering functional brain dynamics. In particular, we proposed to extract the time points showing significant alignment or liberality with the structural brain scaffold through comparison with surrogate data. Compared to the majority of dynamic functional connectivity works, which rely on the successive computation of second-order statistics (e.g., Pearson's correlation coefficient) to quantify the evolution of relationships between brain regions~\cite{preti1701p}, the GSP framework permits to remain at a frame-wise temporal resolution level. Further, as we have also shown above, it harmoniously generalizes to extended settings, such as a joint spatial/temporal decomposition or a localized decomposition of functional signals.

We would like to emphasize that the GSP approach offers a highly flexible framework to analyze functional imaging datasets, where analysis can be conducted on either functional or structural connectivity, and either on a graph that describes the average connectivity across all subjects, or on one graph per subject. The \textit{bimodal} component of the approach, where the constructed graph is used to study functional brain signals, can actually be seen as a double-edged sword: on the one hand, additional information (e.g., structural connectivity) can inform the understanding of functional brain signals, but on the other hand, the obtained results are then strongly dependent on the accuracy of the graph representation itself, and necessitate an underlying relationship between the graph used and the brain signals on top of it.

A number of intriguing connections of GSP with other approaches could be explored. For instance, the GSP methodology allows one to incorporate models of linear diffusion by selecting the spectral window function $g(\lambda)$ as the so-called \textit{diffusion kernel}~\cite{Newman.2010}. Therefore, graph filtering can correspond to diffusion operations of graph signals on the structural graph. A diffusion kernel puts large weights to low-frequency modes (i.e., structurally aligned in our terminology) and decreasing weights as the frequencies increase (i.e., anatomically liberal). Such a network diffusion model on a structural graph has already been used to model disease progression in dementia~\cite{Raj.2012} or to relate structural graphs to functional ones~\cite{Abdelnour.2014}. The link with computational and simulation-based neuroscience is another topic for future interest~\cite{Schirner.2015}; e.g., how eigenmodes capture neural field theory predictions~\cite{Robinson.2016}. 

There is also a clear tendency to refine the granularity of the brain graphs, either by considering finer parcellation schemes~\cite{Atasoy.2016}, or by using voxel-wise approaches through explicit~\cite{behjat1501} or implicit~\cite{preti1702p} representations of the adjacency matrix. The availability of large data from neuroimaging initiatives such as the Human Connectome Project \cite{HCP_paper} has contributed significantly to establishing these refined representations.

%Thus, analyzing the alignment of measured correlations on anatomical networks may provide an important extension to understand signals in the time domain in the human connectome. Mathematically, the same phenomenon for functional as well as anatomical networks implies that the eigenspectrum of the anatomical network is not that different from functional connectivity. Speculatively, this could provide a meaningful and efficient method to reliably recover anatomical networks from functional connectivity networks.

%%%%%%%%%%%%%%%%%%%%%%%%%%%%%%%%%%%%%%%%%%%%%%%%%%%%%%%%%%%%%%%%%%%%%
%%%   B   I   B   L   I   O   %%%%%%%%%%%%%%%%%%%%%%%%%%%%%%%%%%%
%%%%%%%%%%%%%%%%%%%%%%%%%%%%%%%%%%%%%%%%%%%%%%%%%%%%%%%%%%%%%%%%%%%%%
%
\urlstyle{same}
%\bibliographystyle{IEEEtran}
%\bibliography{citations,biblio,slepian}

\end{document}